\documentclass[article,12pt,sort]{elsarticle}

\usepackage{amssymb}
\usepackage{amsmath}
\usepackage{amsthm}

\usepackage{booktabs}
\usepackage{multirow}
\usepackage{subcaption}
\usepackage{caption}
\usepackage{graphicx}
\usepackage{adjustbox}
\usepackage{array}

\newtheorem{definition}{Definition}
\newenvironment{customdefinition}[1]
  {\renewcommand\thedefinition{#1}%
   \definition}
  {\enddefinition}

\usepackage{amsfonts}
\usepackage{graphicx}
\usepackage{textcomp}

\usepackage{bm}

\usepackage{amsthm}
\usepackage{lipsum} 
\usepackage{amsmath}
\usepackage{longtable, threeparttablex}
\usepackage{url}

\usepackage{booktabs}
\usepackage{colortbl}
\usepackage{xcolor}

\usepackage{color}

\usepackage[normalem]{ulem}
\useunder{\uline}{\ul}{}
\usepackage{graphicx}
\usepackage{subcaption}

\usepackage{supertabular}
\usepackage{array}

\usepackage{multicol,lipsum}

\newcolumntype{C}{>{\centering\arraybackslash}p{\dimexpr(\columnwidth-10\tabcolsep)/5\relax}}

\makeatletter
\def\ps@pprintTitle{%
 \let\@oddhead\@empty
 \let\@evenhead\@empty
 \def\@oddfoot{}%
 \let\@evenfoot\@oddfoot}
\makeatother

\usepackage{fancyhdr} 
\usepackage{hyperref}
\begin{document}

\begin{frontmatter}



\title{A Graph Neural Network-based Model with Out-of-Distribution Robustness for Enhancing Antiretroviral Therapy Outcome Prediction for HIV-1}

\author[1,2]{{Giulia} {Di Teodoro}\corref{contrib}} 
\ead{(Corresponding author) diteodoro@diag.uniroma1.it}
\author[1]{Federico Siciliano\corref{contrib}}
\ead{siciliano@diag.uniroma1.it}
\author[3]{Valerio Guarrasi\corref{contrib}}
\ead{valerio.guarrasi@unicampus.it}
\author[5,6]{Anne-Mieke Vandamme}
\ead{annemie.vandamme@kuleuven.be}
\author[7]{Valeria Ghisetti}
\ead{valeria.ghisetti@unito.it}
\author[8,9]{Anders Sönnerborg}
\ead{anders.sonnerborg@ki.se}
\author[4]{Maurizio Zazzi}
\ead{maurizio.zazzi@unisi.it}
\author[1]{Fabrizio Silvestri}
\ead{fsilvestri@diag.uniroma1.it}
\author[1]{Laura Palagi}
\ead{palagi@diag.uniroma1.it}

\cortext[contrib]{These authors contributed equally to the paper}

\affiliation[1]{organization={Sapienza University of Rome, Department of Computer Control and Management Engineering Antonio Ruberti}, postcode = {00185}, city={Rome}, country={Italy}}
\affiliation[2]{organization={EuResist Network}, postcode = {00152}, city={Rome}, country={Italy}}
\affiliation[3]{organization={Unit of Computer Systems and Bioinformatics, Department of Engineering, University Campus Bio-Medico of Rome},postcode={00128}, city ={Rome}, country={Italy}}
\affiliation[4]{organization = {Department of Medical Biotechnologies, University of Siena}, postcode = {53100}, city = {Siena}, country = {Italy}}
\affiliation[5]{organization = {KU Leuven, Department of Microbiology, Immunology and Transplantation, Rega Institute for Medical Research, Clinical and Epidemiological Virology},  city = {Leuven}, country = {Belgium}}
\affiliation[6]{organization = {Center for Global Health and Tropical Medicine, Instituto de Higiene e Medicina Tropical, Universidade Nova de Lisboa}, postcode = {1349-008}, city = {Lisbon}, country = {Portugal}}
\affiliation[7]{organization = {Molecular Biology and Microbiology Unit, Amedeo di Savoia Hospital, ASL Città di Torino}, postcode={10128}, city={Turin}, country={Italy}}
\affiliation[8]{organization ={Karolinska Institutet, Division of Infectious Diseases, Department of Medicine Huddinge}, postcode ={14152}, city={Stockholm}, country={Sweden}}
\affiliation[9]{organization={Karolinska University Hospital, Department of Infectious Diseases},postcode={14186}, city ={Stockholm}, country={Sweden}}

\begin{abstract}
Predicting the outcome of antiretroviral therapies (ART) for HIV-1 is a pressing clinical challenge, especially when the ART includes drugs with limited effectiveness data. This scarcity of data can arise either due to the introduction of a new drug to the market or due to limited use in clinical settings, resulting in clinical dataset with highly unbalanced therapy representation.
To tackle this issue, we introduce a novel joint fusion model, which combines features from a Fully Connected (FC) Neural Network and a Graph Neural Network (GNN) in a multi-modality fashion. Our model uses both tabular data about genetic sequences and a knowledge base derived from Stanford drug-resistance mutation tables, which serve as benchmark references for deducing in-vivo treatment efficacy based on the viral genetic sequence. By leveraging this knowledge base structured as a graph, the GNN component enables our model to adapt to imbalanced data distributions and account for Out-of-Distribution (OoD) drugs.
We evaluated these models' robustness against OoD drugs in the test set.
Our comprehensive analysis demonstrates that the proposed model consistently outperforms the FC model.
These results underscore the advantage of integrating Stanford scores in the model, thereby enhancing its generalizability and robustness, but also extending its utility in contributing in more informed clinical decisions with limited data availability.
The source code is available at \url{https://github.com/federicosiciliano/graph-ood-
hiv}.
\end{abstract}

\begin{keyword}
Graph Neural Network \sep Human Immunodeficiency Virus \sep Knowledge Base \sep Out-of-Distribution \sep Stanford Score \sep Therapy Prediction
\end{keyword}

\end{frontmatter}

\textcopyright{ 2025. Licensed under the Creative Commons  \href{https://creativecommons.org/licenses/by-nc-nd/4.0/}{CC-BY-NC-ND 4.0}.}
\section{Introduction}\label{sec:sample1}

Human immunodeficiency virus (HIV) is an infectious agent that attacks the immune system, and if not treated properly, it can lead to severe consequences. It was discovered in 1981 and thereafter it has infected more than 80 million people and caused the deaths of more than 40 million. By the end of 2021, the World Health Organization estimated that, globally, about 38.4 million people were living with HIV~\cite{WHO}. There are no vaccines for HIV, and it cannot be cured yet. Despite this, with regular treatment and high-quality health care, people living with HIV (PLHIV) can have a life expectancy comparable to that of people without the virus. 

Antiretroviral therapy (ART) has been shown to be effective in controlling viral replication for most PLHIV. If no ART is used, HIV-infected patients come to manifest symptoms of acquired immunodeficiency syndrome (AIDS). Commonly, an ART usually consists of a cocktail of three antiretroviral drugs that cooperatively block virus replication at multiple steps and,  compared to a single drug, are successful in prolonging the time it takes for HIV to become resistant to therapy. This success is due to the development of potent and well-tolerated antiretroviral drugs with high genetic barriers to resistance. 
However, before a change of therapy, patient records should be carefully analyzed to guide treatment choices and preserve future treatment options. This includes reviewing past drug exposure, treatment failures, cumulative drug resistance, and duration of virological suppression~\cite{nihGuidelinesAntiretroviral,eacsocietyEACSGuidelines}. 

By examining the correlation between HIV-1 mutational patterns and drug resistance, it is possible to interpret a genotype and provide a model of drug susceptibility. There are two types of genotype interpretation systems (GISs): rules-based systems and systems based on statistical or Machine Learning (ML) models. The former is based on rules derived from the knowledge of experts who have created tables of drug-mutation resistance~\cite{Wensing2022-od,Anderson2008-tt,Frentz2010-ic,Rhee22,vanDerKlundert_2022}. These tables guide the assignment of resistance scores to specific genotypes. The main rule sets from ANRS, HIVdb~\cite{Tang2012-po}, HIV-GRADE and Rega Institute are accessible on the HIV-GRADE~\cite{Obermeier2012-pu} platform. Drug resistance scores and consequently rules-based approaches are updated frequently to reflect new findings on HIV drug resistance based on expert consensus~\cite{Paredes2017-gm}. 
The latter is trained directly on genotypic data for predicting therapy outcomes~\cite{Beerenwinkel2003-dj,Lengauer2006-zz,Pironti2017}.
Early iterations of data-driven GIS, such as VircoTYPE's VirtualPhenotype$^{\textit{TM}}$~\cite{Vermeiren2007-xp} and the geno2pheno framework~\cite{Beerenwinkel2003-dj}, aimed to predict the phenotype \textit{in-vitro}. VircoTYPE initially calculated the phenotypic effect by pooling the impacts of individual mutations on individual drugs via a linear regression algorithm, later modified to provide clinically actionable resistance estimates~\cite{Winters2008-nj}. Geno2pheno, on the other hand, originally served to estimate single-drug resistance and was later extended to predict 3-4-drug ART outcomes. Geno2pheno-THEO used both genotypic information and user-provided data to predict the probability of treatment success~\cite{Altmann2009-vu}. Geno2pheno[resistance]\cite{Pironti2017-wi}\cite{Lengauer2006-zz} makes use of a support vector machine model to estimate specific drug resistance, while Geno2pheno[drug exposure]\cite{Pironti2017} first calculates drug exposure scores and then uses these scores in a second model used for predicting treatment outcome. In a recent approach \cite{Di_Teodoro2023-nr}, a linear-Support Vector Machine model is used for predicting therapy outcome, having as input the drug combinations and all the mutations detected in the previous genotypic tests, i.e. incorporating historical information.

With the availability of increasing amounts of genotypic and clinical data, ML models have become widely used to guide treatment choices, especially for treatment-experienced patients with complicated drug-resistance profiles. When training ML models to predict treatment outcomes ~\cite{guarrasi2023multi, caruso2022multimodal, caruso2023deep, Raparelli2023-xw, guarrasi2022optimized, guarrasi2024multimodal, rofena2024deep, ruffini2024multi, guarrasi2024systematic}, certain treatment regimens may be omitted from the training set due to limited data on specific drugs. This creates an Out of Distribution (OoD) problem, where models encounter data that significantly deviates from their training set, potentially leading to performance issues or incorrect predictions. This issue often arises if a drug is rarely used in regimens or is newly introduced, resulting in a lack of accessible efficacy data. As treatment strategies for HIV patients have evolved over time, there is a significant imbalance in how different therapies are represented in available clinical datasets. Consequently, while some treatments are well-represented with numerous examples, others are scarcely documented, with only a few samples available. 
In this scenario, classical ML models and existing statistical methods commonly used for predicting outcomes of HIV therapies cannot be used to predict the outcome of drug regimens containing OoD drugs. At the same time, rules-based systems may be inaccurate when assigning a susceptibility score to a therapy containing a new drug because of limited clinical experience with that drug.  This can result in a preliminary resistance score for the new drug, which may require adjustments as more data becomes available. In the clinical setting, there is a substantial benefit for treatment decision support tools informing use of OoD drugs. The most intuitive case is the need to employ with confidence a novel drug in individuals who are not able any more to control virus replication and hence at risk of clinical deterioration. When a novel drug is first licensed, available resistance data are often limited to in-vitro studies, such as site-directed mutagenesis and susceptibility testing on laboratory or clinical isolates~\cite{Wensing2022-od}. This challenge is further compounded when pivotal clinical trials intentionally exclude participants with known resistance issues. By enrolling only patients on first-line therapy without pre-existing mutations, there is often no established relationship between mutations and the in vivo efficacy of the drug. Lack of knowledge or under-appreciation of resistance pathways may lead to unwise use of novel drugs, further compromising treatment options in heavily treatment experienced subjects. On the other hand, old drugs with limited use in the past may come back to be useful in certain circumstances, including constraints derived from tolerability issues and drug-drug interactions which are quite common in the elderly population living with HIV. For these reasons, predicting the effectiveness of therapies with an OoD drug would be highly valuable to safely exploit the full complement of antiretroviral therapy in a case-by-case scenario.   \\
All previous methods for predicting the ARTs outcomes typically incorporate the viral genotype and the constituent drugs of the therapy as input variables, lacking models for the impact of each specific drug within the therapy on the treatment's outcome. Furthermore, such models fail to tackle the issue of sparse and uneven representation of ARTs within the training dataset. Prior researches \cite{bogojeska_bioinformatics,Bogojeska2012-wj} addressed the issue of unbalanced therapy representation in clinical datasets by employing separate models for each therapy combination or for each individual drugs and assuming that individual drugs in each ART have additive effects on the therapy response, but that is not always the case. They fail to address scenarios where ARTs include OoD drugs in the test set. Additionally, these methods necessitate training multiple models rather than a single, unified model.
Until now, the literature mainly focused on OoD detection with several methods. Early approaches used predictive uncertainty thresholds ~\cite{hendrycks2016baseline} and temperature scaling ~\cite{liang2017enhancing}. Deep ensembles were later introduced for better uncertainty estimation ~\cite{lakshminarayanan2017simple}. Recent advancements include self-supervision for richer feature extraction ~\cite{hendrycks_2019, jihoon_NEURIPS2020} and transfer learning, which fine-tunes classifiers on large datasets to improve OoD detection accuracy ~\cite{cohen2022, liznerski2022exposing}. However, research focusing on OoD in HIV treatment prediction is scarce.\\
To overcome the presented limitations derived from uneven and sparce representation of therapies and OoD drugs in the test set, we propose a novel joint fusion model that uses Fully Connected (FC) and Graph Neural Networks (GNNs) \cite{Scarselli2009-vo} which are able to incorporate knowledge derived from the drug-resistance mutation tables~\cite{Wensing2022-cr}. In this way, we can accurately predict the outcome of therapies that contain drugs for which no data were present in the training phase but are present in the knowledge base. Additionally, it allows to adapt the model to imbalanced therapies data distribution, accounting for non-linear drug interactions, and to train a single model in a end-to-end fashion. Moreover, our research aims to diverge from the conventional approach of solely focusing on OoD detection: rather than identifying and flagging OoD instances, our focus is to cultivate model robustness against OoD scenarios in the realm of HIV therapy prediction. Our efforts contribute to the development of more reliable and effective predictive models for HIV prognosis, ultimately leading to improved patient care and treatment decision-making.
 The primary contributions of our work are detailed as follows: 
\begin{enumerate}
    \item We introduce a novel joint fusion model combining FC Neural Networks and GNNs to predict the outcome of ART for HIV-1.
    \item The proposed model leverages both tabular data and structured knowledge bases, incorporating drug-resistance mutation scores from the Stanford HIVdb to enhance treatment predictions.
    \item Our approach addresses the challenge of OoD drugs, improving model robustness in scenarios where drugs in the test set are not represented in the training data.
    \item By capturing non-linear interactions between drugs and mutations, the model overcomes limitations of existing approaches that assume additive drug effects in ART combinations.
    \item Unlike traditional models, our fusion model can adapt to imbalanced data distributions and allows training a single, unified model, enhancing generalization capabilities in clinical settings.
    \item Experimental results demonstrate that the proposed model significantly outperforms standalone models, especially in scenarios involving OoD drugs, thus improving clinical decision support for complex HIV cases.
\end{enumerate}

In the following, 
in section~\ref{sec:dataset} we present the data used by the methodology presented in section~\ref{sec:methodology}. Next, in section~\ref{sec:experiments} we present the different experiments implemented to obtain the results presented and discussed in section~\ref{sec:results}. Finally, in section~\ref{sec:conclusions} we conclude by highlighting the main findings opening future work for the following research.

\section{Dataset}\label{sec:dataset}
The Euresist Integrated Database (EIDB)~\cite{Rossetti} is one of the largest databases containing data on PLHIV, both treatment-naïve and treatment-experienced, who have been followed since 1998. Established in 2006, the EIDB collects information from nine national cohorts: Italy, Germany, Sweden, Portugal, Spain, Luxembourg, Belgium, Turkey, and Russia. The data, in anonymized form, cover demographic and clinical aspects of PLHIV, such as types of ARTs, reasons for therapy changes, treatment responses, CD4+ cell counts, viral load (VL) measurements, and additional viral co-infections.
The database includes data from $105101$ PLHIV, but the information from many patients is not consistent enough for our analysis. The database has been analyzed to select consistent and complete data to be part of the dataset used in the experiments. 

The model we propose is therapy-oriented and therefore the data are structured according to therapy-patient pairs. In the analysis of treatment success or failure, the concepts of the patient-treatment episode (PTE) is delineated, as mentioned in previous research~\cite{Zazzi2011-pg}.
\begin{customdefinition}{PTE}
This episode is a collection of data for a patient to assess the response to a treatment. It consists of a genotype (protease (PR) reverse transcriptase (RT) and/or integrase (IN)) at baseline, the group of pharmacological compounds used in antiretroviral treatment (cART), an optional VL at baseline, taken no more than 90 days before the start of treatment, and follow-up VLs. PTEs include both first-line therapy and treatment change episodes referred to a patient.
\end{customdefinition}

The dataset we built contains information coming from PTEs. The outcome of drug treatment is indicated with a label, $y \in \{0,1\}$, representing success or failure. This is based on a recently adopted definition of the EuResist standard datum.

\begin{customdefinition}{Standard datum} To determine the success of a therapy, a follow-up VL, and optionally a VL at the baseline, is needed. All the follow-up VLs measured between $20$ and $28$ weeks after therapy initiations are taken. Among them, the VL whose measurement day is closer to the $24^{th}$ week is considered. The PTE is labeled as a success if this follow-up VL is less than $50$ copies of HIV-1 RNA per milliliter of blood plasma. Alternatively, the treatment for that patient is considered a failure. \\
For cases where treatment was modified before reaching the $24$-week, the following criteria are used:
\begin{itemize}
    \item Therapies that last $4$ weeks at maximum are not included because they are most likely terminated due to adverse effects.
    \item Therapies that last from $4$ to $8$ weeks are considered a success if the last detected VL is under $50$ copies/ml or shows at least $1$ log reduction compared to the baseline VL; otherwise, they are considered a failure.
    \item Therapies that last from $8$ to $20$ weeks are considered a success if the last detected 
    VL is under $50$ copies/ml or shows at least $2$ log reduction compared to the baseline VL; otherwise, they are considered a failure.
\end{itemize} 
\end{customdefinition}

To be included in the dataset, a patient-therapy pair must fulfill the following requirements:
\begin{itemize}
    \item Conform to the definition of PTE, including details of the compounds used, at least one genotypic sequence before the start of therapy and the follow-up VL.
    \item The success or failure of the patient's therapy must be able to be determined in accordance with the definition of Standard Datum.
\end{itemize}
The drugs included in the dataset are listed: lamivudine (3TC), abacavir (ABC), amprenavir (APV), atazanavir (ATV), zidovudine (AZT), bictegravir (BIC), cabotegravir (CAB), stavudine (D4T), zalcitabine (DDC), didanosine (DDI), delavirdine (DLV), doravirine (DOR), darunavir (DRV), dolutegravir (DTG), efavirenz (EFV), etravirine (ETR), elvitegravir (EVG), fosamprenavir (FPV), emtricitabine (FTC), indinavir (IDV), lopinavir (LPV), nelfinavir (NFV), nevirapine (NVP), raltegravir (RAL), rilpivirine (RPV), saquinavir (SQV), tenofovir alafenamide (TAF), tenofovir disoproxil (TDF), tipranavir (TPV).\\
Both polymorphic and non-polymorphic~\cite{Shafer2007-qr} mutations are considered, with the rationale that unrecognized mutations or combinations could influence the virus's susceptibility to cART or its fitness. \\ 
Ultimately, the dataset consists of 22000 patient-therapy pairs: 12386 successes and 9614 failures.

\section{Methodology}
\label{sec:methodology}

\subsection{Setting}
\label{sec:setting}

Let's define \( m_j \) as a particular mutation, with \( m_j \in M = \{m_1, m_2, \ldots, m_{N_M}\}\), where \( M \) contains all the mutations considered. \( d_k \) represents an individual drug, \( d_k \in D = \{d_1, d_2, \ldots, d_{N_D}\} \), comprising the drugs under examination. \( N_M \) represents the total number of mutations in consideration, while \( N_D \) denotes the total number of drugs being studied.

\begin{figure*}[t]
  \centering
  \includegraphics[width=\linewidth]{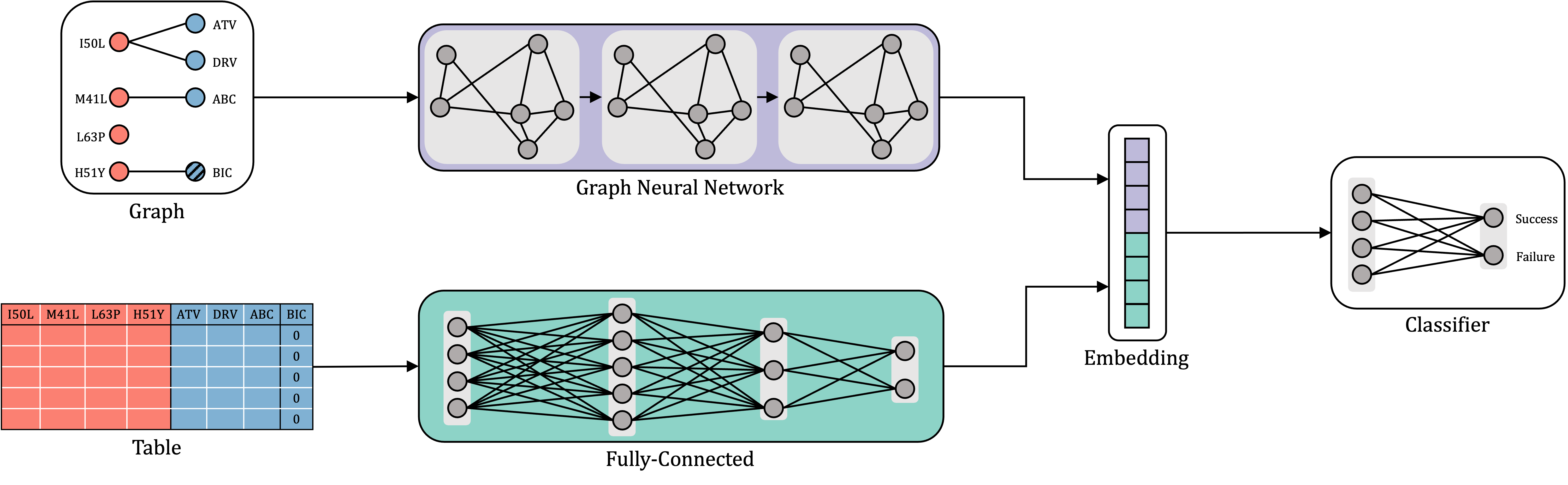}
  \caption{\textbf{Schematic view of the pipeline.}}
  \label{fig:pipeline}
\end{figure*}

We introduce a binary vector \( r \in \{0, 1\}^{N_M} \), which flags the presence of drug-resistant mutations (DRMs) that have been observed in the last available viral genotype prior to initiating the antiretroviral treatment of interest.
Additionally, \( z \in \{0, 1\}^{N_D} \) is a binary vector that specifies the combination of drugs employed in the target ART for which we want to predict the outcome.
A binary label \( y \in \{0, 1\} \) categorizes the treatment as either successful (\( y = 0 \)) or unsuccessful (\( y = 1 \)), based on criteria outlined in the Standard Datum definition given in section \ref{sec:dataset}.
The model we propose uses the dataset both in tabular form and with data graph structures, as it can be seen in Figure \ref{fig:pipeline}. Here below, it is explained how these two data types are treated.

\subsection{Tabular data} \label{subsec: tabular_data}
The tabular data consists of a set of records in which each data point corresponds to a unique patient-treatment pair \( x_i = (r_i, z_i) \), along with the associated outcome \( y_i \). Thus, each data point can be represented as a triplet \( (r_i, z_i, y_i) \).
The training set \( T = \{(r_1, z_1, y_1), \ldots, (r_N, z_N, y_N)\} \) is assembled, where \( N \) is the total number of patient-treatment pairs.

\subsection{Graph data} \label{subsec: graph_data}
There exist a relationship $s_{m_jd_k} \in [-1,+1]$ between a mutation $m_j \in M$ and a drug $d_k \in D$.
These drug-mutation relationships are shown in the Stanford tables and are regarded in HIVdb as genotypic resistance scores~\cite{Tang2012-po}. The score associated with a drug-mutation pair $s_{m_jd_k}$, which we will refer to as the Stanford score, is higher the more a mutation decreases the susceptibility of the virus to a drug.

For each tabular record \( (x_i, y_i) = (r_i, z_i, y_i)\), we generate a graph structure, where each patient-drug sample \( (x_i,y_i) \) is represented by a graph. Nodes in the graph represent both mutations and drugs. The edges
are directional and
connect a mutation node $n_{m_j}$ to a drug node $n_{d_k}$ if a Stanford score $s_{m_jd_k}$ is associated with that mutation-drug pair; nodes not involved in any mutation-drug pair and for which a Stanford score exists remain isolated nodes.
Each node, $n_{d_k}$ or $n_{m_j}$, in the graph has a binary attribute $r_{ij}$ or $z_{ik}$, respectively, valued at 1 if the corresponding mutation $m_j$ or drug $d_k$ is present in the record $r_i$ or $z_i$, respectively, 0 otherwise. Each edge $e_{m_jd_k}$ in the graph connecting a mutation $m_j$ to a drug $d_k$ is associated with a weight equal to the Stanford score value $s_{m_jd_k}$ associated with that drug-mutation pair.

To further refine our representation, we created two different types of graphs for each record in the training set \(T\): 
\begin{itemize}
    \item Type ``all'': the graph $G_i$ related to a record \( (x_i,y_i) \) includes nodes for all mutations and drugs in \(T\).
    \item Type ``subset'': the graph $G_i$ related to a record \( (x_i,y_i) \) includes only the nodes for the mutations and drugs that are actually present in \(x_i\), i.e., have attribute value 1.
\end{itemize}
Below, we will add a suffix to the model name to differentiate models trained on different types of graphs. The suffix added is ``$_A$'' or ``$_S$'' depending on whether the model is trained on ``all'' or ``subset'' type graphs, respectively.

\subsection{Datasets to simulate OoD features}
The OoD features for which we want to test the prediction capabilities of the proposed model are drugs, to align with the clinical need to predict the outcome of therapies that include drugs for which there are insufficient training data. This lack of data may result from the fact that the drug is new to the market or because currently available data are incomplete or insufficient.
To simulate scenarios in which a feature is not observed in the training set but is present in the test set as an OoD, we generate as many different datasets as there are OoD features on which we want to test the model. In each of these datasets, the training set consists only and exclusively of samples in which the OoD drug has value 0. This means that the training set will contain all and only ARTs that do not involve that drug. In contrast, the test set is composed only of samples in which the OoD drug is present and valued at 1. The test set therefore will consist of all and only the ARTs that contain that drug.

\subsection{Models}
\subsubsection{Fully Connected Neural Networks}
\label{sec:fully_connected_neural_networks}
The Fully Connected (FC) neural network takes as input a feature vector of size 5970, which is the sum of the number of mutations (\(N_M = 5941\)) and drugs (\(N_D = 29\)). The FC network architecture consists of two hidden layers with increasing sizes (64 and 256). Each hidden layer applies a linear transformation followed by a Rectified Linear Unit (ReLU) activation function. The output layer performs a linear transformation to produce the final output with a size equal to the number of output classes (2).

\subsubsection{Graph Neural Networks}
The central idea behind GNNs is to update the representation of a node by aggregating information from its neighbors.
This aggregation procedure captures the local structure of the graph and can be iteratively applied to capture information from further away in the graph.
A typical GNN layer involves several interconnected stages. 

The adopted GNN layer~\cite{you2020design} incorporates a linear layer followed by several components: \textit{(i)} batch normalization $\text{BN}(\cdot)$, \textit{(ii)} a nonlinear activation function $\text{ACT}(\cdot)$, and \textit{(iii)} an aggregation function $\text{AGG}(\cdot)$. The formal representation of the $l$-th GNN layer is:

\[
h^{(l+1)}_i = \text{AGG}\left(\left\{\text{ACT}\left(\text{BN}\left(W^{(l)}h^{(l)}_u + b^{(l)}\right)\right), u \in \mathcal{N}(i)\right\}\right),
\]

where $h_i$ denotes the embedding of node $i$ at the $l$-th layer, $W^{(l)}$ and $b^{(l)}$ are the learnable parameters, and $\mathcal{N}(i)$ represents the neighborhood of node $i$.

The proposed GNN's architecture consists of three hidden layers with increasing sizes (4, 8, and 16), ReLU activation functions and the mean as aggregation function.
Each hidden layer uses the message-passing layers operation with additive multi-head attention (5 heads) and directed message-passing.
Then a Global Max Pooling operation is performed creating a feature vector, on which the output layer performs a linear transformation to produce the final output with a size equal to the number of output classes (2).

\subsubsection{MIX - Joint Fusion}\label{sec:mix}
The architecture of the fusion module combines the features extracted from the Graph Neural Network (GNN) and the FC networks, as illustrated in Figure \ref{fig:pipeline}. For this fusion, the features from the last hidden layers of both networks are concatenated into an embedding. This embedding is then passed through a linear layer, followed by a sigmoid activation function, producing a final output vector with a size equal to the number of output classes (2). The concatenation of features allows the model to capture complementary information from both networks and make more informed predictions.

\subsection{Training}
The training phase involves training the GNN and the FC network independently. Then, utilizing transfer learning, the fused networks are trained in an end-to-end manner, leveraging the knowledge gained from the individual network training stages. This approach allows the fused networks to benefit from the specific features learned by each network while also capturing synergistic effects that arise from their combination. The fused networks can learn higher-level representations that effectively leverage the strengths of both architectures. This transfer learning strategy enhances the overall performance and enables the fused networks to make more accurate predictions compared to training them from scratch.
During training, the GNN, FC, and fusion modules are trained using the Cross-Entropy Loss, which measures the difference between predicted and actual class labels. To prevent bias, the Cross-Entropy Loss is weighted based on the class distribution. The optimizer used for training is Adam with a learning rate of 0.001. The models are trained for 1000 epochs, iterating over the training dataset multiple times to optimize the network parameters and improve performance.

\section{Experiments}
\label{sec:experiments}
This section explains how experiments were conducted to test the MIX model's effectiveness in improving generalization abilities, in the case of OoD features in the test set, compared to an FC neural network. \\
For each drug, we trained a model without data on that drug in the training set. Then we tested the model with a test set with solely records using that drug to test the model's ability in the case of OoD. 

We did not run experiments for drugs for which the procedure of dividing the data between training and test sets would have resulted in too few samples in the test set or even all data points with the same class label (all successes or all failures). In these cases, the size and quality of the test set were not adequate to obtain meaningful and reliable results. \\
 For each model, the dataset was partitioned in a consistent manner into a training set, validation set, and test set with 60\%, 20\%, and 20\% of the data, respectively. The training and validation sets remain fixed across models, while the test set varies by including out-of-distribution instances corresponding to the out-of-distribution feature being evaluated. Care was taken to ensure that treatments related to an individual patient were grouped into the same subset.  This measure was instituted to mitigate the risk that the model might inadvertently learn patterns from patients' medical records if their treatments were scattered between the training and testing sets. Such a situation could compromise the model by skewing predictive results for future treatments given to that same patient. By consolidating treatments from the same patient into a single data subset, we aim to attenuate any undue bias and ensure that the model's predictive capacity is grounded in unbiased generalization as opposed to individual medical histories.  Further details about the prevention of data leakage can be found in Appendix A. \\
We compare the performances of the following models: 
\begin{itemize}
    \item FC: Fully connected neural network trained on tabular data presented in subsection \ref{subsec: tabular_data}.
    \item GNN$_A$: Graph neural network trained on samples represented in a type ``all'' graphs, as explained in subsection \ref{subsec: graph_data}.
    \item GNN$_S$: Graph neural network trained on samples represented in a type ``subset'' graphs, as explained in subsection \ref{subsec: graph_data}.
    \item MIX$_A$: Fusion model that combines a GNN trained on type ``all'' graphs and an FC neural network trained on tabular data.
    \item MIX$_S$: Fusion model that combines a GNN trained on type ``subset'' graphs and an FC neural network trained on tabular data.
\end{itemize}

The performances of GNN and MIX models are compared with those of FC by performing a paired Wilcoxon test.

 The source code is available at: \url{https://anonymous.4open.science/r/hiv-gnn/}.


\section{Results}
\label{sec:results}

Results are shown in Table \ref{tab:results_performance}  and plotted in Table A.1 of the Appendix.
We reported the results of Accuracy (Acc), of the Receiving Operating Characteristics (ROC) area under the curve and of the precision-recall (PR) area under the curve in percentages. The choice to report these metrics is due to the fact that test sets with an OoD feature can be unbalanced by presenting more successes than failures. The rows represent the model type and the OoD column indicates which is the OoD drug in the test set. \\
If we look at the ROC score results more in detail, we see that the only case in which FC is better than the other models it is when the OoD feature is ETR. In this case, the differences in performance between the models are not significant. This is indicative of the fact that knowledge derived from Stanford scores in this case does not help in better prediction.
ETR has been the most elusive drug in terms of prediction of in-vivo effectiveness based on HIV genotype. There are three main reasons for this. First, contrary to most other drugs, ETR susceptibility is impacted by many mutations, amplifying mutation interaction effects which are difficult to examine both in-vitro and in-vivo~\cite{Vingerhoets2010-jn,Kagan2009-er}. Indeed, the interpretation of susceptibility to ETR has been changed several times and alternative scores have concomitantly existed for a long time~\cite{Vingerhoets2012-nt}. Second, due to substantial toxicity and inconvenience of dosing (twice daily instead of once daily), ETR has been used sparingly, generating few data for training genotype interpretation systems. Genotype interpretation has remained based on the original DUET trials where ETR was combined with the highly potent DRV, hence the effects of the individual drug were largely confounded by the accompanying drug~\cite{Vingerhoets2010-jn}. Third, drug toxicity and inconvenience of dosing per sè increased the risk of treatment failure as a result of poor tolerability and low adherence to therapy rather than of true virological non-efficacy.

\captionsetup[subtable]{labelformat=empty}

\begin{table*}[htbp]
\centering
\caption{Test set performances (\%). The best and second-best results are in \textbf{bold} and \underline{underlined}, respectively. * The Wilcoxon test is significant at level 0.01, $\star$ at level 0.05.}
\label{tab:results_performance}

\setlength{\tabcolsep}{3pt} 
\renewcommand{\arraystretch}{0.9} 

\begin{tabular}{cccc}
\begin{subtable}[t]{0.23\textwidth}
\centering
\caption{ATV}
\scriptsize 
\adjustbox{max width=\linewidth}{
\begin{tabular}{lccc}
\toprule
\textbf{Model} & \textbf{Acc} & \textbf{ROC} & \textbf{PR} \\
\midrule
FC & \begin{tabular}[c]{@{}l@{}} 56.7 \\ ($\pm0.89$)  \end{tabular} & \begin{tabular}[c]{@{}l@{}} 66.2\\ ($\pm0.95$)  \end{tabular} & \begin{tabular}[c]{@{}l@{}} 62.7\\ ($\pm1.38$)  \end{tabular} \\
GNN$_A$ & \begin{tabular}[c]{@{}l@{}} 61.5*\\ ($\pm0.90$)  \end{tabular} & \begin{tabular}[c]{@{}l@{}} 64.6\\ ($\pm0.86$)  \end{tabular} & \begin{tabular}[c]{@{}l@{}} 59.9\\ ($\pm1.47$)  \end{tabular} \\
GNN$_S$ & \begin{tabular}[c]{@{}l@{}} \textbf{62.1}*\\ ($\pm0.78$)  \end{tabular} & \begin{tabular}[c]{@{}l@{}} 64.2\\ ($\pm0.99$)  \end{tabular} & \begin{tabular}[c]{@{}l@{}} 60.3\\ ($\pm1.43$)  \end{tabular} \\
MIX$_A$ & \begin{tabular}[c]{@{}l@{}} \underline{61.9}*\\ ($\pm0.82$)  \end{tabular} & \begin{tabular}[c]{@{}l@{}} \textbf{66.8}\\ ($\pm0.93$)  \end{tabular} & \begin{tabular}[c]{@{}l@{}} \textbf{64.4}\\ ($\pm1.21$)  \end{tabular} \\
MIX$_S$ & \begin{tabular}[c]{@{}l@{}} 61.8*\\ ($\pm0.84$)  \end{tabular} & \begin{tabular}[c]{@{}l@{}} \underline{66.7}\\ ($\pm0.95$)  \end{tabular} & \begin{tabular}[c]{@{}l@{}} \underline{64.3}\\ ($\pm1.15$)  \end{tabular} \\
\bottomrule
\end{tabular}
}
\end{subtable}
&
\begin{subtable}[t]{0.23\textwidth}
\centering
\caption{DRV}
\scriptsize
\adjustbox{max width=\linewidth}{
\begin{tabular}{lccc}
\toprule
\textbf{Model} & \textbf{Acc} & \textbf{ROC} & \textbf{PR} \\
\midrule
FC & \begin{tabular}[c]{@{}l@{}} 58.8\\ ($\pm0.70$)  \end{tabular} & \begin{tabular}[c]{@{}l@{}} 59.7\\ ($\pm1.04$)  \end{tabular} & \begin{tabular}[c]{@{}l@{}} \underline{46.5}\\ ($\pm1.44$)  \end{tabular} \\
GNN$_A$ & \begin{tabular}[c]{@{}l@{}} 60.4$\star$\\ ($\pm0.76$)  \end{tabular} & \begin{tabular}[c]{@{}l@{}} 60.2\\ ($\pm0.93$)  \end{tabular} & \begin{tabular}[c]{@{}l@{}} 44.5 \\ ($\pm1.57$)  \end{tabular} \\
GNN$_S$ & \begin{tabular}[c]{@{}l@{}} 63.7*\\ ($\pm0.67$)  \end{tabular} & \begin{tabular}[c]{@{}l@{}} \underline{60.5}\\ ($\pm0.99$)  \end{tabular} & \begin{tabular}[c]{@{}l@{}}44.8 \\ ($\pm1.56$)  \end{tabular} \\
MIX$_A$ & \begin{tabular}[c]{@{}l@{}} \textbf{64.8}*\\ ($\pm0.74$)  \end{tabular} & \begin{tabular}[c]{@{}l@{}} \textbf{61.0}  \\ ($\pm0.95$)  \end{tabular}& \begin{tabular}[c]{@{}l@{}} \textbf{46.6}\\ ($\pm1.30$)  \end{tabular} \\
MIX$_S$ & \begin{tabular}[c]{@{}l@{}}\underline{64.5}* \\ ($\pm0.78$)  \end{tabular} & \begin{tabular}[c]{@{}l@{}} \textbf{61.0}\\ ($\pm0.97$)  \end{tabular} & \begin{tabular}[c]{@{}l@{}} \textbf{46.6}\\ ($\pm1.40$)  \end{tabular} \\
\bottomrule
\end{tabular}
}
\end{subtable}
&
\begin{subtable}[t]{0.23\textwidth}
\centering
\caption{DTG}
\scriptsize
\adjustbox{max width=\linewidth}{
\begin{tabular}{lccc}
\toprule
\textbf{Model} & \textbf{Acc} & \textbf{ROC} & \textbf{PR} \\
\midrule
FC & \begin{tabular}[c]{@{}l@{}} 75.1 \\ ($\pm0.82$)  \end{tabular} & \begin{tabular}[c]{@{}l@{}} 46.4 \\ ($\pm1.75$)  \end{tabular} & \begin{tabular}[c]{@{}l@{}} 15.0 \\ ($\pm1.24$)  \end{tabular} \\
GNN$_A$ & \begin{tabular}[c]{@{}l@{}} \underline{75.8} \\ ($\pm0.90$)  \end{tabular} & \begin{tabular}[c]{@{}l@{}} 48.3 \\ ($\pm1.73$)  \end{tabular} & \begin{tabular}[c]{@{}l@{}} 15.2\\ ($\pm1.20$)  \end{tabular} \\
GNN$_S$ & \begin{tabular}[c]{@{}l@{}} \textbf{82.7}* \\ ($\pm0.82$)  \end{tabular} & \begin{tabular}[c]{@{}l@{}} \textbf{51.5}\\ ($\pm1.59$)  \end{tabular} & \begin{tabular}[c]{@{}l@{}} \textbf{16.2}\\ ($\pm1.10$)  \end{tabular} \\
MIX$_A$ & \begin{tabular}[c]{@{}l@{}} 72.0 \\ ($\pm0.90$)  \end{tabular} & \begin{tabular}[c]{@{}l@{}} \underline{49.0}\\ ($\pm1.52$)  \end{tabular} & \begin{tabular}[c]{@{}l@{}} \underline{16.0} \\ ($\pm1.15$)  \end{tabular}\\
MIX$_S$ & \begin{tabular}[c]{@{}l@{}} 72.0 \\ ($\pm0.92$)  \end{tabular}& \begin{tabular}[c]{@{}l@{}} \underline{49.0}\\ ($\pm1.55$)  \end{tabular} & \begin{tabular}[c]{@{}l@{}} \underline{16.0}\\ ($\pm1.27$)  \end{tabular} \\
\bottomrule
\end{tabular}
}
\end{subtable}
&
\begin{subtable}[t]{0.23\textwidth}
\centering
\caption{EFV}
\scriptsize
\adjustbox{max width=\linewidth}{
\begin{tabular}{lccc}
\toprule
\textbf{Model} & \textbf{Acc} & \textbf{ROC} & \textbf{PR} \\
\midrule
FC & \begin{tabular}[c]{@{}l@{}} 64.5 \\ ($\pm0.99$)  \end{tabular} & \begin{tabular}[c]{@{}l@{}} \underline{67.7} \\ ($\pm0.99$)  \end{tabular} & \begin{tabular}[c]{@{}l@{}} \underline{61.3} \\ ($\pm1.65$)  \end{tabular} \\
GNN$_A$ & \begin{tabular}[c]{@{}l@{}} 62.7  \\ ($\pm0.81$)  \end{tabular} & \begin{tabular}[c]{@{}l@{}}65.4 \\ ($\pm1.04$)  \end{tabular}  & \begin{tabular}[c]{@{}l@{}} 60.1 \\ ($\pm1.64$)  \end{tabular} \\
GNN$_S$ & \begin{tabular}[c]{@{}l@{}} 62.7\\ ($\pm1.04$)  \end{tabular} & \begin{tabular}[c]{@{}l@{}} 65.9 \\ ($\pm1.06$)  \end{tabular} & \begin{tabular}[c]{@{}l@{}} 60.8  \\ ($\pm1.52$)  \end{tabular} \\
MIX$_A$ & \begin{tabular}[c]{@{}l@{}} \underline{64.8}\\ ($\pm0.87$)  \end{tabular} & \begin{tabular}[c]{@{}l@{}} \textbf{69.2}\\ ($\pm0.96$)  \end{tabular} & \begin{tabular}[c]{@{}l@{}} \textbf{65.0}\\ ($\pm1.32$)  \end{tabular} \\
MIX$_S$ & \begin{tabular}[c]{@{}l@{}} \textbf{65.0}\\ ($\pm0.92$)  \end{tabular} & \begin{tabular}[c]{@{}l@{}}\textbf{69.2}\\ ($\pm1.09$)  \end{tabular} & \begin{tabular}[c]{@{}l@{}} \textbf{65.0}\\ ($\pm1.58$)  \end{tabular} \\
\bottomrule
\end{tabular}
}
\end{subtable} \\[-2mm] 

\begin{subtable}[t]{0.23\textwidth}
\centering
\caption{ETR}
\scriptsize
\adjustbox{max width=\linewidth}{
\begin{tabular}{lccc}
\toprule
\textbf{Model} & \textbf{Acc} & \textbf{ROC} & \textbf{PR} \\
\midrule
FC & \begin{tabular}[c]{@{}l@{}} 60.8\\ ($\pm1.65$)  \end{tabular} & \begin{tabular}[c]{@{}l@{}}  \textbf{60.8}\\ ($\pm2.27$)  \end{tabular} &\begin{tabular}[c]{@{}l@{}} \underline{49.6}\\ ($\pm3.26$)  \end{tabular}  \\
GNN$_A$ & \begin{tabular}[c]{@{}l@{}} \underline{61.7}\\ ($\pm1.55$)  \end{tabular} & \begin{tabular}[c]{@{}l@{}}59.3 \\ ($\pm2.17$)  \end{tabular} & \begin{tabular}[c]{@{}l@{}} 45.0\\ ($\pm3.14$)  \end{tabular} \\
GNN$_S$ & \begin{tabular}[c]{@{}l@{}} 58.7\\ ($\pm1.64$)  \end{tabular} & \begin{tabular}[c]{@{}l@{}} 56.7\\ ($\pm2.17$)  \end{tabular} & \begin{tabular}[c]{@{}l@{}} 43.8\\ ($\pm2.94$)  \end{tabular} \\
MIX$_A$ & \begin{tabular}[c]{@{}l@{}} \textbf{62.4}\\ ($\pm1.74$)  \end{tabular} & \begin{tabular}[c]{@{}l@{}} \underline{60.6}\\ ($\pm2.11$)  \end{tabular} & \begin{tabular}[c]{@{}l@{}}47.4 \\ ($\pm3.24$)  \end{tabular} \\
MIX$_S$ & \begin{tabular}[c]{@{}l@{}} 59.6\\ ($\pm1.76$)  \end{tabular} & \begin{tabular}[c]{@{}l@{}} 59.9\\ ($\pm2.27$)  \end{tabular} & \begin{tabular}[c]{@{}l@{}} \textbf{49.8} \\ ($\pm3.01$)  \end{tabular} \\
\bottomrule
\end{tabular}
}
\end{subtable}
&
\begin{subtable}[t]{0.23\textwidth}
\centering
\caption{EVG}
\scriptsize
\adjustbox{max width=\linewidth}{
\begin{tabular}{lccc}
\toprule
\textbf{Model} & \textbf{Acc} & \textbf{ROC} & \textbf{PR} \\
\midrule
FC & \begin{tabular}[c]{@{}l@{}} 74.4\\ ($\pm1.51$)  \end{tabular} & \begin{tabular}[c]{@{}l@{}} 56.0\\ ($\pm3.29$)  \end{tabular} & \begin{tabular}[c]{@{}l@{}} 18.3 \\ ($\pm2.35$)  \end{tabular}\\
GNN$_A$ & \begin{tabular}[c]{@{}l@{}} \textbf{83.2}*\\ ($\pm1.32$)  \end{tabular} & \begin{tabular}[c]{@{}l@{}} 56.4\\ ($\pm2.86$)  \end{tabular} & \begin{tabular}[c]{@{}l@{}}18.6 \\ ($\pm2.52$)  \end{tabular} \\
GNN$_S$ & \begin{tabular}[c]{@{}l@{}} 53.6\\ ($\pm1.91$)  \end{tabular} & \begin{tabular}[c]{@{}l@{}} 56.6 \\ ($\pm3.08$)  \end{tabular}& \begin{tabular}[c]{@{}l@{}} 19.1\\ ($\pm2.53$)  \end{tabular} \\
MIX$_A$ & \begin{tabular}[c]{@{}l@{}} 72.1\\ ($\pm1.66$)  \end{tabular} & \begin{tabular}[c]{@{}l@{}} \underline{57.3}\\ ($\pm2.79$)  \end{tabular} & \begin{tabular}[c]{@{}l@{}} \underline{20.7} \\ ($\pm3.26$)  \end{tabular}\\
MIX$_S$ & \begin{tabular}[c]{@{}l@{}} \underline{74.9}\\ ($\pm1.57$)  \end{tabular} & \begin{tabular}[c]{@{}l@{}} \textbf{58.0}\\ ($\pm3.14$)  \end{tabular} & \begin{tabular}[c]{@{}l@{}} \textbf{21.9} \\ ($\pm3.19$)  \end{tabular} \\
\bottomrule
\end{tabular}
}
\end{subtable}
&
\begin{subtable}[t]{0.23\textwidth}
\centering
\caption{FPV}
\scriptsize
\adjustbox{max width=\linewidth}{
\begin{tabular}{lccc}
\toprule
\textbf{Model} & \textbf{Acc} & \textbf{ROC} & \textbf{PR} \\
\midrule
FC & \begin{tabular}[c]{@{}l@{}} 64.4 \\ ($\pm1.80$)  \end{tabular} & \begin{tabular}[c]{@{}l@{}}\underline{64.5} \\ ($\pm2.51$)  \end{tabular} & \begin{tabular}[c]{@{}l@{}}75.1 \\ ($\pm2.52$)  \end{tabular} \\
GNN$_A$ & \begin{tabular}[c]{@{}l@{}} 64.1\\ ($\pm1.95$)  \end{tabular} & \begin{tabular}[c]{@{}l@{}} 63.0\\ ($\pm2.14$)  \end{tabular} & \begin{tabular}[c]{@{}l@{}} \underline{75.7}\\ ($\pm2.32$)  \end{tabular} \\
GNN$_S$ & \begin{tabular}[c]{@{}l@{}} 64.1 \\ ($\pm1.76$)  \end{tabular}& \begin{tabular}[c]{@{}l@{}} 63.2\\ ($\pm2.14$)  \end{tabular} & \begin{tabular}[c]{@{}l@{}} 75.5\\ ($\pm2.55$)  \end{tabular} \\
MIX$_A$ & \begin{tabular}[c]{@{}l@{}} \textbf{69.0}*\\ ($\pm1.91$)  \end{tabular} & \begin{tabular}[c]{@{}l@{}}\textbf{67.6} \\ ($\pm2.27$)  \end{tabular} & \begin{tabular}[c]{@{}l@{}} \textbf{77.2} \\ ($\pm2.55$)  \end{tabular}\\
MIX$_S$ & \begin{tabular}[c]{@{}l@{}}\underline{68.9}* \\ ($\pm1.65$)  \end{tabular} & \begin{tabular}[c]{@{}l@{}} \textbf{67.6}\\ ($\pm2.26$)  \end{tabular} & \begin{tabular}[c]{@{}l@{}} \textbf{77.2} \\ ($\pm2.65$)  \end{tabular}\\
\bottomrule
\end{tabular}
}
\end{subtable}
&
\begin{subtable}[t]{0.23\textwidth}
\centering
\caption{IDV}
\scriptsize
\adjustbox{max width=\linewidth}{
\begin{tabular}{lccc}
\toprule
\textbf{Model} & \textbf{Acc} & \textbf{ROC} & \textbf{PR} \\
\midrule
FC & \begin{tabular}[c]{@{}l@{}} 67.7\\ ($\pm2.21$)  \end{tabular} & \begin{tabular}[c]{@{}l@{}} 62.7\\ ($\pm2.99$)  \end{tabular} & \begin{tabular}[c]{@{}l@{}} \underline{77.9}\\ ($\pm2.61$)  \end{tabular} \\
GNN$_A$ & \begin{tabular}[c]{@{}l@{}} 70.2*\\ ($\pm2.01$)  \end{tabular} & \begin{tabular}[c]{@{}l@{}} \textbf{64.5}\\ ($\pm2.79$)  \end{tabular} & \begin{tabular}[c]{@{}l@{}} 77.7 \\ ($\pm2.57$)  \end{tabular}\\
GNN$_S$ & \begin{tabular}[c]{@{}l@{}} 70.4$\star$\\ ($\pm2.11$)  \end{tabular} & \begin{tabular}[c]{@{}l@{}}\textbf{64.5} \\ ($\pm2.42$)  \end{tabular} & \begin{tabular}[c]{@{}l@{}} \textbf{78.3}\\ ($\pm2.87$)  \end{tabular} \\
MIX$_A$ & \begin{tabular}[c]{@{}l@{}} \textbf{71.1}*\\ ($\pm2.26$)  \end{tabular} & \begin{tabular}[c]{@{}l@{}} \underline{63.9}\\ ($\pm3.08$)  \end{tabular} & \begin{tabular}[c]{@{}l@{}} 76.9\\ ($\pm2.77$)  \end{tabular} \\
MIX$_S$ & \begin{tabular}[c]{@{}l@{}} \underline{70.9}$\star$\\ ($\pm1.98$)  \end{tabular} & \begin{tabular}[c]{@{}l@{}}63.7 \\ ($\pm2.59$)  \end{tabular}& \begin{tabular}[c]{@{}l@{}} 76.6\\ ($\pm2.71$)  \end{tabular} \\
\bottomrule
\end{tabular}
}
\end{subtable} \\[-2mm] 

\begin{subtable}[t]{0.23\textwidth}
\centering
\caption{LPV}
\scriptsize
\adjustbox{max width=\linewidth}{
\begin{tabular}{lccc}
\toprule
\textbf{Model} & \textbf{Acc} & \textbf{ROC} & \textbf{PR} \\
\midrule
FC & \begin{tabular}[c]{@{}l@{}} 62.8\\ ($\pm0.79$)  \end{tabular} & \begin{tabular}[c]{@{}l@{}} 66.5\\ ($\pm0.91$)  \end{tabular} & \begin{tabular}[c]{@{}l@{}} \underline{73.4}\\ ($\pm1.12$)  \end{tabular} \\
GNN$_A$ & \begin{tabular}[c]{@{}l@{}} 61.8\\ ($\pm0.86$)  \end{tabular} & \begin{tabular}[c]{@{}l@{}} 64.5 \\ ($\pm0.89$)  \end{tabular}& \begin{tabular}[c]{@{}l@{}} 72.0 \\ ($\pm1.21$)  \end{tabular}\\
GNN$_S$ & \begin{tabular}[c]{@{}l@{}} \underline{63.2}\\ ($\pm0.83$)  \end{tabular} & \begin{tabular}[c]{@{}l@{}} 64.0\\ ($\pm1.04$)  \end{tabular} & \begin{tabular}[c]{@{}l@{}} 71.5\\ ($\pm1.09$)  \end{tabular} \\
MIX$_A$ & \begin{tabular}[c]{@{}l@{}} \textbf{65.2}*\\ ($\pm0.86$)  \end{tabular} & \begin{tabular}[c]{@{}l@{}} \textbf{67.4}\\ ($\pm0.80$)  \end{tabular} & \begin{tabular}[c]{@{}l@{}} \textbf{73.8}\\ ($\pm1.21$)  \end{tabular} \\
MIX$_S$ & \begin{tabular}[c]{@{}l@{}} \textbf{65.2}*\\ ($\pm0.79$)  \end{tabular} & \begin{tabular}[c]{@{}l@{}} \underline{67.3}\\ ($\pm1.05$)  \end{tabular} & \begin{tabular}[c]{@{}l@{}} \textbf{73.8}\\ ($\pm1.23$)  \end{tabular} \\
\bottomrule
\end{tabular}
}
\end{subtable}
&
\begin{subtable}[t]{0.23\textwidth}
\centering
\caption{NFV}
\scriptsize
\adjustbox{max width=\linewidth}{
\begin{tabular}{lccc}
\toprule
\textbf{Model} & \textbf{Acc} & \textbf{ROC} & \textbf{PR} \\
\midrule
FC & \begin{tabular}[c]{@{}l@{}} \underline{65.3}\\ ($\pm1.63$)  \end{tabular} & \begin{tabular}[c]{@{}l@{}} 62.5\\ ($\pm2.34$)  \end{tabular} & \begin{tabular}[c]{@{}l@{}} 74.0\\ ($\pm2.39$)  \end{tabular} \\
GNN$_A$ & \begin{tabular}[c]{@{}l@{}} 63.4\\ ($\pm1.66$)  \end{tabular} & 6\begin{tabular}[c]{@{}l@{}} 1.4\\ ($\pm2.20$)  \end{tabular} & \begin{tabular}[c]{@{}l@{}}73.0 \\ ($\pm2.45$)  \end{tabular} \\
GNN$_S$ & \begin{tabular}[c]{@{}l@{}} 63.5\\ ($\pm1.69$)  \end{tabular} & \begin{tabular}[c]{@{}l@{}} \underline{63.6}\\ ($\pm1.98$)  \end{tabular} & \begin{tabular}[c]{@{}l@{}} \underline{74.7}\\ ($\pm2.31$)  \end{tabular} \\
MIX$_A$ & \begin{tabular}[c]{@{}l@{}} \textbf{67.2}*\\ ($\pm1.75$)  \end{tabular} & \begin{tabular}[c]{@{}l@{}} \textbf{67.0} \\ ($\pm2.17$)  \end{tabular}& \begin{tabular}[c]{@{}l@{}} \textbf{77.9}\\ ($\pm1.95$)  \end{tabular} \\
MIX$_S$ & \begin{tabular}[c]{@{}l@{}} \textbf{67.2}*\\ ($\pm1.75$)  \end{tabular} & \begin{tabular}[c]{@{}l@{}} \textbf{67.0}\\ ($\pm1.85$)  \end{tabular} & \begin{tabular}[c]{@{}l@{}} \textbf{77.9}\\ ($\pm1.84$)  \end{tabular} \\
\bottomrule
\end{tabular}
}
\end{subtable}
&
\begin{subtable}[t]{0.23\textwidth}
\centering
\caption{NVP}
\scriptsize
\adjustbox{max width=\linewidth}{
\begin{tabular}{lccc}
\toprule
\textbf{Model} & \textbf{Acc} & \textbf{ROC} & \textbf{PR} \\
\midrule
FC & \begin{tabular}[c]{@{}l@{}} 63.8\\ ($\pm1.23$)  \end{tabular} & \begin{tabular}[c]{@{}l@{}}73.2 \\ ($\pm1.20$)  \end{tabular} & \begin{tabular}[c]{@{}l@{}} \textbf{73.0}\\ ($\pm1.98$)  \end{tabular} \\
GNN$_A$ & \begin{tabular}[c]{@{}l@{}} 60.6\\ ($\pm1.43$)  \end{tabular} & \begin{tabular}[c]{@{}l@{}} 69.0\\ ($\pm1.62$)  \end{tabular} & \begin{tabular}[c]{@{}l@{}} 71.3\\ ($\pm1.81$)  \end{tabular} \\
GNN$_S$ & \begin{tabular}[c]{@{}l@{}} 56.3\\ ($\pm1.30$)  \end{tabular} & \begin{tabular}[c]{@{}l@{}} 69.1\\ ($\pm1.39$)  \end{tabular} & \begin{tabular}[c]{@{}l@{}} 71.0\\ ($\pm1.97$)  \end{tabular} \\
MIX$_A$ & \begin{tabular}[c]{@{}l@{}} \underline{69.9}*\\ ($\pm1.24$)  \end{tabular} & \begin{tabular}[c]{@{}l@{}} \underline{73.8}\\ ($\pm1.37$)  \end{tabular} & \begin{tabular}[c]{@{}l@{}} 72.6\\ ($\pm1.85$)  \end{tabular} \\
MIX$_S$ & \begin{tabular}[c]{@{}l@{}} \textbf{70.4}*\\ ($\pm1.18$)  \end{tabular} & \begin{tabular}[c]{@{}l@{}} \textbf{74.1}\\ ($\pm1.27$)  \end{tabular} & \begin{tabular}[c]{@{}l@{}} \underline{72.8}\\ ($\pm1.96$)  \end{tabular} \\
\bottomrule
\end{tabular}
}
\end{subtable}
&
\begin{subtable}[t]{0.23\textwidth}
\centering
\caption{RAL}
\scriptsize
\adjustbox{max width=\linewidth}{
\begin{tabular}{lccc}
\toprule
\textbf{Model} & \textbf{Acc} & \textbf{ROC} & \textbf{PR} \\
\midrule
FC & \begin{tabular}[c]{@{}l@{}} \textbf{61.6}\\ ($\pm0.94$)  \end{tabular} & \begin{tabular}[c]{@{}l@{}} 56.9\\ ($\pm1.06$)  \end{tabular} & \begin{tabular}[c]{@{}l@{}} 39.2\\ ($\pm1.79$)  \end{tabular} \\
GNN$_A$ & \begin{tabular}[c]{@{}l@{}} 49.8\\ ($\pm1.12$)  \end{tabular} & \begin{tabular}[c]{@{}l@{}} 57.6\\ ($\pm1.30$)  \end{tabular} & \begin{tabular}[c]{@{}l@{}}38.7 \\ ($\pm1.75$)  \end{tabular} \\
GNN$_S$ & \begin{tabular}[c]{@{}l@{}} 54.1\\ ($\pm1.18$)  \end{tabular} & \begin{tabular}[c]{@{}l@{}} 55.7\\ ($\pm1.27$)  \end{tabular} & \begin{tabular}[c]{@{}l@{}}36.7 \\ ($\pm1.72$)  \end{tabular} \\
MIX$_A$ & \begin{tabular}[c]{@{}l@{}} 54.1\\ ($\pm1.05$)  \end{tabular} & \begin{tabular}[c]{@{}l@{}} \textbf{60.5}\\ ($\pm1.38$)  \end{tabular} & \begin{tabular}[c]{@{}l@{}} \textbf{43.4}\\ ($\pm1.68$)  \end{tabular} \\
MIX$_S$ & \begin{tabular}[c]{@{}l@{}} \underline{61.1}\\ ($\pm1.09$)  \end{tabular} & \begin{tabular}[c]{@{}l@{}} \underline{60.0}\\ ($\pm1.29$)  \end{tabular} & \begin{tabular}[c]{@{}l@{}} \underline{41.1}\\ ($\pm2.02$)  \end{tabular} \\
\bottomrule
\end{tabular}
}
\end{subtable} \\[-2mm] 

\begin{subtable}[t]{0.23\textwidth}
\centering
\caption{RPV}
\scriptsize
\adjustbox{max width=\linewidth}{
\begin{tabular}{lccc}
\toprule
\textbf{Model} & \textbf{Acc} & \textbf{ROC} & \textbf{PR} \\
\midrule
FC & \begin{tabular}[c]{@{}l@{}} 54.9\\ ($\pm1.48$)  \end{tabular} & \begin{tabular}[c]{@{}l@{}}57.5 \\ ($\pm3.40$)  \end{tabular} & \begin{tabular}[c]{@{}l@{}}\textbf{15.8} \\ ($\pm3.06$)  \end{tabular} \\
GNN$_A$ & \begin{tabular}[c]{@{}l@{}} \textbf{83.5}*\\ ($\pm1.15$)  \end{tabular} & \begin{tabular}[c]{@{}l@{}} \textbf{61.6}\\ ($\pm3.11$)  \end{tabular} & \begin{tabular}[c]{@{}l@{}} 14.2 \\ ($\pm2.94$)  \end{tabular}\\
GNN$_S$ & \begin{tabular}[c]{@{}l@{}} \underline{82.5}*\\ ($\pm1.13$)  \end{tabular} & \begin{tabular}[c]{@{}l@{}} \underline{59.2}\\ ($\pm3.23$)  \end{tabular} & \begin{tabular}[c]{@{}l@{}} 14.2\\ ($\pm2.80$)  \end{tabular} \\
MIX$_A$ & \begin{tabular}[c]{@{}l@{}} 81.3*\\ ($\pm1.27$)  \end{tabular} & \begin{tabular}[c]{@{}l@{}} 56.9\\ ($\pm3.02$)  \end{tabular} & \begin{tabular}[c]{@{}l@{}} \textbf{15.8}\\ ($\pm3.61$)  \end{tabular} \\
MIX$_S$ & \begin{tabular}[c]{@{}l@{}} 81.0*\\ ($\pm1.10$)  \end{tabular} & \begin{tabular}[c]{@{}l@{}} 56.9\\ ($\pm3.33$)  \end{tabular} & \begin{tabular}[c]{@{}l@{}} \underline{15.6}\\ ($\pm3.29$)  \end{tabular} \\
\bottomrule
\end{tabular}
}
\end{subtable}
&
\begin{subtable}[t]{0.23\textwidth}
\centering
\caption{SQV}
\scriptsize
\adjustbox{max width=\linewidth}{
\begin{tabular}{lccc}
\toprule
\textbf{Model} & \textbf{Acc} & \textbf{ROC} & \textbf{PR} \\
\midrule
FC & \begin{tabular}[c]{@{}l@{}}67.7 \\ ($\pm1.60$)  \end{tabular} & \begin{tabular}[c]{@{}l@{}} \underline{60.5}\\ ($\pm2.04$)  \end{tabular} & \begin{tabular}[c]{@{}l@{}} \textbf{77.8}\\ ($\pm2.08$)  \end{tabular} \\
GNN$_A$ & \begin{tabular}[c]{@{}l@{}} 65.6\\ ($\pm1.85$)  \end{tabular} & \begin{tabular}[c]{@{}l@{}} 60.2\\ ($\pm2.13$)  \end{tabular} & \begin{tabular}[c]{@{}l@{}} 75.4\\ ($\pm2.01$)  \end{tabular} \\
GNN$_S$ & \begin{tabular}[c]{@{}l@{}} \underline{68.8}\\ ($\pm1.61$)  \end{tabular} & \begin{tabular}[c]{@{}l@{}} 59.8\\ ($\pm2.28$)  \end{tabular} & \begin{tabular}[c]{@{}l@{}}75.7 \\ ($\pm2.20$)  \end{tabular} \\
MIX$_A$ & \begin{tabular}[c]{@{}l@{}} \textbf{69.7}$\star$\\ ($\pm1.57$)  \end{tabular} & \begin{tabular}[c]{@{}l@{}} \textbf{61.2}\\ ($\pm2.20$)  \end{tabular} & \begin{tabular}[c]{@{}l@{}} \underline{76.6}\\ ($\pm2.40$)  \end{tabular} \\
MIX$_S$ & \begin{tabular}[c]{@{}l@{}} \textbf{69.7}$\star$\\ ($\pm1.60$)  \end{tabular} & \begin{tabular}[c]{@{}l@{}} \textbf{61.2}\\ ($\pm2.13$)  \end{tabular} & \begin{tabular}[c]{@{}l@{}} \underline{76.6}\\ ($\pm2.19$)  \end{tabular} \\
\bottomrule
\end{tabular}
}
\end{subtable}
&
\begin{subtable}[t]{0.23\textwidth}
\centering
\caption{TPV}
\scriptsize
\adjustbox{max width=\linewidth}{
\begin{tabular}{lccc}
\toprule
\textbf{Model} & \textbf{Acc} & \textbf{ROC} & \textbf{PR} \\
\midrule
FC & \begin{tabular}[c]{@{}l@{}} 75.4\\ ($\pm1.99$)  \end{tabular} & \begin{tabular}[c]{@{}l@{}} 59.1\\ ($\pm3.81$)  \end{tabular} & \begin{tabular}[c]{@{}l@{}} \textbf{79.7}\\ ($\pm2.94$)  \end{tabular} \\
GNN$_A$ & \begin{tabular}[c]{@{}l@{}} \textbf{75.7}\\ ($\pm2.51$)  \end{tabular} & \begin{tabular}[c]{@{}l@{}} 52.1\\ ($\pm2.98$)  \end{tabular} & \begin{tabular}[c]{@{}l@{}} \underline{79.2}\\ ($\pm2.74$)  \end{tabular} \\
GNN$_S$ & \begin{tabular}[c]{@{}l@{}} \underline{75.5}\\ ($\pm2.15$)  \end{tabular} & \begin{tabular}[c]{@{}l@{}} 57.3\\ ($\pm4.53$)  \end{tabular} & \begin{tabular}[c]{@{}l@{}} 78.5\\ ($\pm3.05$)  \end{tabular} \\
MIX$_A$ & \begin{tabular}[c]{@{}l@{}} 73.0\\ ($\pm2.64$)  \end{tabular} & \begin{tabular}[c]{@{}l@{}} \textbf{59.3}\\ ($\pm3.80$)  \end{tabular} & \begin{tabular}[c]{@{}l@{}} 78.8\\ ($\pm2.96$)  \end{tabular} \\
MIX$_S$ & \begin{tabular}[c]{@{}l@{}} 73.0\\ ($\pm2.38$)  \end{tabular} & \begin{tabular}[c]{@{}l@{}} \underline{59.2}\\ ($\pm3.52$)  \end{tabular} & \begin{tabular}[c]{@{}l@{}} 78.8\\ ($\pm3.41$)  \end{tabular} \\
\bottomrule
\end{tabular}
}
\end{subtable}
&
\begin{subtable}[t]{0.23\textwidth}
\centering
\caption{No OoD}
\scriptsize
\adjustbox{max width=\linewidth}{
\begin{tabular}{lccc}
\toprule
\textbf{Model} & \textbf{Acc} & \textbf{ROC} & \textbf{PR} \\
\midrule
FC & \begin{tabular}[c]{@{}l@{}} 71.4 \\ ($\pm0.69$)  \end{tabular}& \begin{tabular}[c]{@{}l@{}} 78.8\\ ($\pm0.67$)  \end{tabular} & \begin{tabular}[c]{@{}l@{}} 72.9\\ ($\pm0.99$)  \end{tabular} \\
GNN$_A$ & \begin{tabular}[c]{@{}l@{}} 68.1\\ ($\pm0.66$)  \end{tabular} & \begin{tabular}[c]{@{}l@{}} 72.5\\ ($\pm0.83$)  \end{tabular} & \begin{tabular}[c]{@{}l@{}}65.3  \\ ($\pm1.28$)  \end{tabular}\\
GNN$_S$ & \begin{tabular}[c]{@{}l@{}} 66.3\\ ($\pm0.67$)  \end{tabular} & \begin{tabular}[c]{@{}l@{}} 68.2\\ ($\pm0.88$)  \end{tabular} & \begin{tabular}[c]{@{}l@{}} 60.3\\ ($\pm1.16$)  \end{tabular} \\
MIX$_A$ & \begin{tabular}[c]{@{}l@{}} \underline{76.1}*\\ ($\pm0.60$)  \end{tabular} & \begin{tabular}[c]{@{}l@{}} \underline{82.9} \\ ($\pm0.61$)  \end{tabular}& \begin{tabular}[c]{@{}l@{}} \underline{77.7}\\ ($\pm1.06$)  \end{tabular} \\
MIX$_S$ & \begin{tabular}[c]{@{}l@{}} \textbf{79.9}*\\ ($\pm0.60$)  \end{tabular} & \begin{tabular}[c]{@{}l@{}} \textbf{85.8}\\ ($\pm0.63$)  \end{tabular} & \begin{tabular}[c]{@{}l@{}} \textbf{80.4}\\ ($\pm1.08$)  \end{tabular} \\
\bottomrule
\end{tabular}
}
\end{subtable} \\
\end{tabular}
\end{table*}

In the case of a test set with an OoD feature, the MIX model performs best in 11 out of 15 in the ROC score and in 9 out of 15 for the PR score. The FC is better than the other models in only 1 case when we look at the ROC score and in 3 when we look at the PR. Furthermore, a statistically significant difference exists between the MIX model and the FC model in 9 out of 15 cases. This shows that the proposed model improves generalization ability compared with a FC.\\
Moreover, the MIX model we propose outperforms the other models when there are no OoD features in the test set with a gap in performances noticeably larger and a statistically significant difference with respect to the FC. This might suggest that the MIX models are leveraging information from both FC and GNN architectures more effectively when there is no OoD data to contend with.\\

Looking at the PR score more in detail, we see that FC is better than the other models only when the OoD feature is TPV, SQV or NVP.\\
Considerations similar to the ones made for ETR certainly apply to TPV, the most potent protease inhibitor with very high genetic barrier, a highly complex resistance mutation profile and perhaps the highest toxicity ever experienced with an antiretroviral drug~\cite{Vergani2011-ox}. Indeed, TPV had very limited use in the clinic, again compromising the development of an accurate genotypic interpretation score. \\
Regarding SQV, particularly in the first years of use without any pharmacokinetic enhancer, suffered from very low bioavailability limiting its in-vivo effectiveness~\cite{Collier1996-zt}. This could cause algorithms that rely on Stanford scores to be less reliable for predicting the success of SQV-containing therapies, especially in the years 1996-2001. The model therefore could predict false successes and lower the performance of PR scores.
\\
For NVP, the FC is the model that performs best but the difference with the other models is minimal. As a resistance profile, unlike ETR and TPV, it is simple.  It is worth noting that NVP is part of older therapies that generally worked less given even the other drugs combined with NVP, and thus NPV therapies could have failed more than predicted through rules-based GIS such as those using Stanford scores~\cite{Sungkanuparph2007-ae}. Consequently, even the MIX model is helped little by the knowledge derived from Stanford scores in predicting failures.


\section{Conclusions}
\label{sec:conclusions}
The primary goal of this study was to predict the outcome of antiretroviral therapies for HIV-1, especially in scenarios with OoD drugs. 
ML models are increasingly used to guide treatment choices due to growing genotypic and clinical data. However, these models have limitations when predicting outcomes for treatments containing drugs not in their training set. This is usually because the drug is either rarely used or newly launched, making relevant data scarce. Rules-based systems also struggle with new drugs, as their susceptibility scores often rely on limited clinical experience. To face this issue, we introduced and evaluated a novel model, denoted as MIX model, a joint fusion model of a GNN and a FC neural network. Thanks to the use of graph structures, we incorporated the knowledge derived from the Stanford score in the data. These scores derive from drug-resistance mutation tables created and continually updated by experts to be used in rules-based GIS and take decisions on new HIV treatments to prescribe to PLHIV.  Our results demonstrated the superiority of our MIX model in handling OoD cases, outperforming the standalone FC model in the majority of the scenarios. Moreover, the MIX model shows superior performance compared to other models when the test set is devoid of OoD drugs. The performance gap is not only remarkable but also statistically significant compared to the FC model. This might indicate that the MIX model is more adept at synthesizing information from the FC and GNN frameworks, particularly when no OoD data are present.\\
The limitations of past approaches \cite{bogojeska_bioinformatics,Bogojeska2012-wj}, when scarce data for certain therapies were available, consisted of implicitly assuming that individual drugs comprising each combination therapy have an additive effect and that more models must be trained to have a prediction system for the ART outcome. In contrast, our model, using GNNs, does not consider drugs only in the specific combinations of therapies but gives the possibility to consider the interaction that each drug has with all other drugs and mutations and makes it possible to model nonlinear relationships between drug effects. Moreover, our approach requires training a single model in an end-to-end fashion. \\
Our research underscores the value of integrating multiple data representations and neural network architectures. It establishes a foundation for improved generalizability, especially in contexts where there's limited data availability or novel data challenges.
In summary, by focusing on fortifying model robustness against OoD scenarios, we pave the way for advancements that transcend traditional OoD detection methodologies, fostering a new paradigm in the field of HIV prediction and medical ML.\\
Moreover, the range of applicability of this model is wider than the HIV field since it can be used in all applications in which a knowledge base is available and can be structured as a knowledge graph. 

In future studies, we will explore further the treatment of drugs that act on pairs of mutations since there are Stanford scores that associate a drug with a pair of mutations that were not used in this study. Instead of traditional graphs that represent binary relationships, a hyper-graph, that allows for modeling higher-order relationships, can be leveraged. 
We will incorporate mutations detected in GRTs older than the one at baseline. Building on a prior study \cite{Di_Teodoro2023-nr}, there exists potential in using all older mutations to improve the generalization of the prediction models. This could allow the model to recognize more nuanced patterns and trends in the data. 

\section*{Author contributions statement}
G.D.T., Fe.S., V.Gu., Fa.S., L.P.: Conceptualization. G.D.T., Fe.S., V.Gu.: Methodology, Software, Investigation, Writing original draft, Review and Editing. G.D.T.: Formal analysis of data. G.D.T., Fe.S.: Project administration. A.M.V., V.Gi., A.S., M.Z.: Data Curation and Revision of the final manuscript. M.Z.: Validation and interpretation of results. Fa.S., L.P.: Supervision.

\section*{Acknowledgments}
The authors would like to thank the EuResist Network working group for their valuable work for the EIDB.
This work was partially supported by projects FAIR (PE0000013) and SERICS (PE00000014) under the MUR National Recovery and Resilience Plan funded by the European Union - NextGenerationEU. Supported also by the ERC Advanced Grant 788893 AMDROMA,  EC H2020RIA project “SoBigData++” (871042), PNRR MUR project  IR0000013-SoBigData.it.

\section*{Data availability statement}
This analysis was conducted using the EIDB. Due to the sensitive nature of personal medical data, it is not feasible to make this data publicly available on the internet. Data were obtained from the EuResist Network and are available for request through a study application form at \url{https://www.euresist.org/become-a-partner} with the permission of the EuResist Network.  Additionally, data from the HIVdb were utilized in this study. The HIVdb is openly accessible at \url{https://hivdb.stanford.edu/ TCEs/}.

\section*{Ethic statement}
Ethical approval was granted in the host countries of the respective original databases contributing data to EIDB.

\bibliographystyle{elsarticle-num-names} 
\bibliography{references_etal}

\clearpage
\appendix

\setcounter{section}{0}
\setcounter{table}{0}
\setcounter{figure}{0}

\section{Details on the datasets and on results}

Given the definitions of Patient-Treatment Episodes (PTEs) that encompass both initial therapy and episodes of treatment changes for a patient, it is possible for multiple records in the datasets, each involving different treatments, to correspond to the same patient receiving various therapies at different times.

Table \ref{tab:therapies per patient} presents the number of therapies administered to the same patient, categorized into successful therapies (label 0) and failed therapies (label 1), as defined by the Standard Datum in Section 2 of the main text.

To prevent data leakage, all therapies related to a particular patient were either entirely included in the training set or entirely in the test set.

\begin{figure*}[]
  \centering
  \begin{subfigure}[b]{\linewidth}
    \centering
    \includegraphics[width=.8\linewidth]{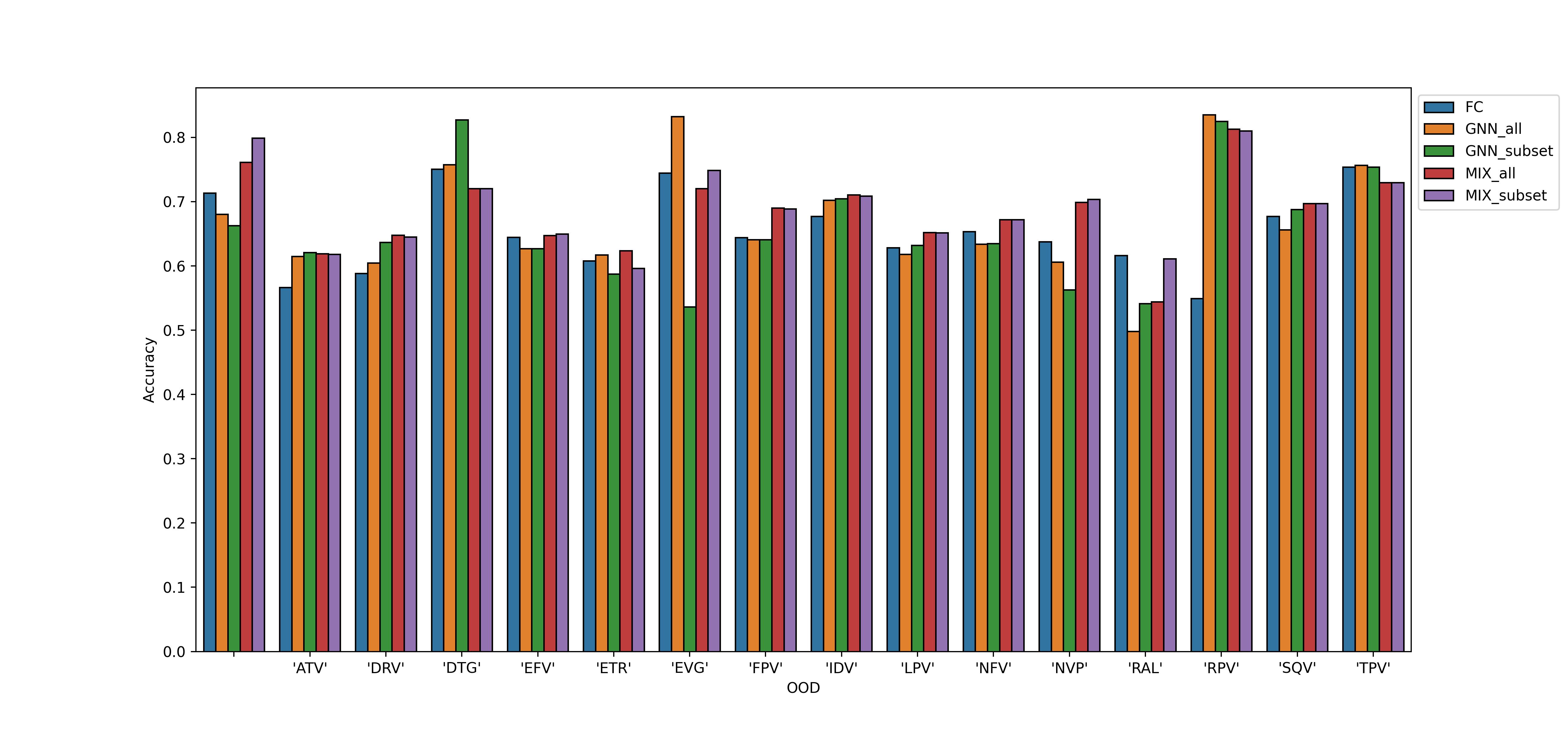}
    \caption{}
    \label{fig:sub1_acc}
  \end{subfigure}
  \hfill
  \begin{subfigure}[b]{\linewidth}
    \centering
    \includegraphics[width=0.8\linewidth]{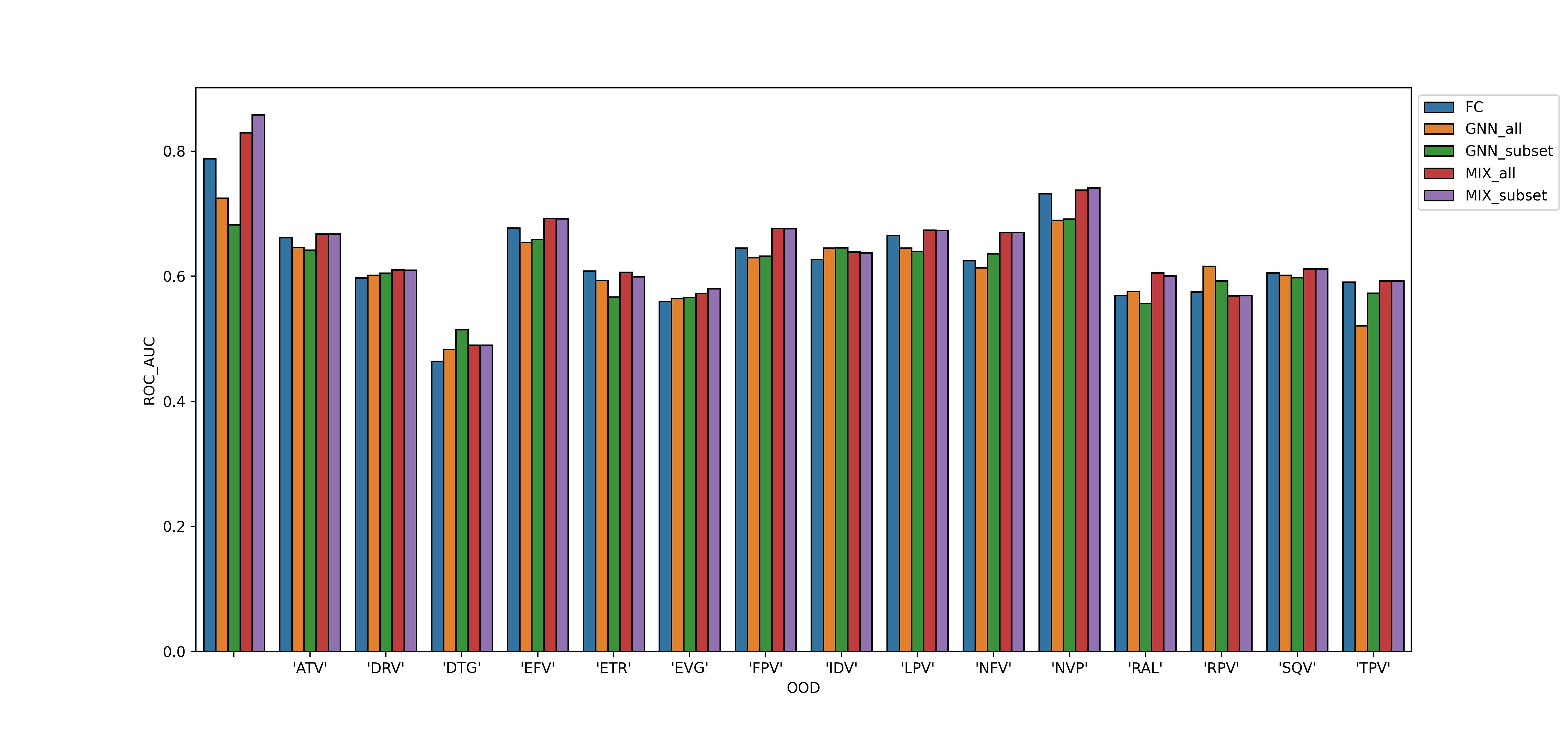}
    \caption{}
    \label{fig:sub1_roc_acu}
  \end{subfigure}
  \hfill
\begin{subfigure}[b]{\linewidth}
    \centering
    \includegraphics[width=.8\linewidth]{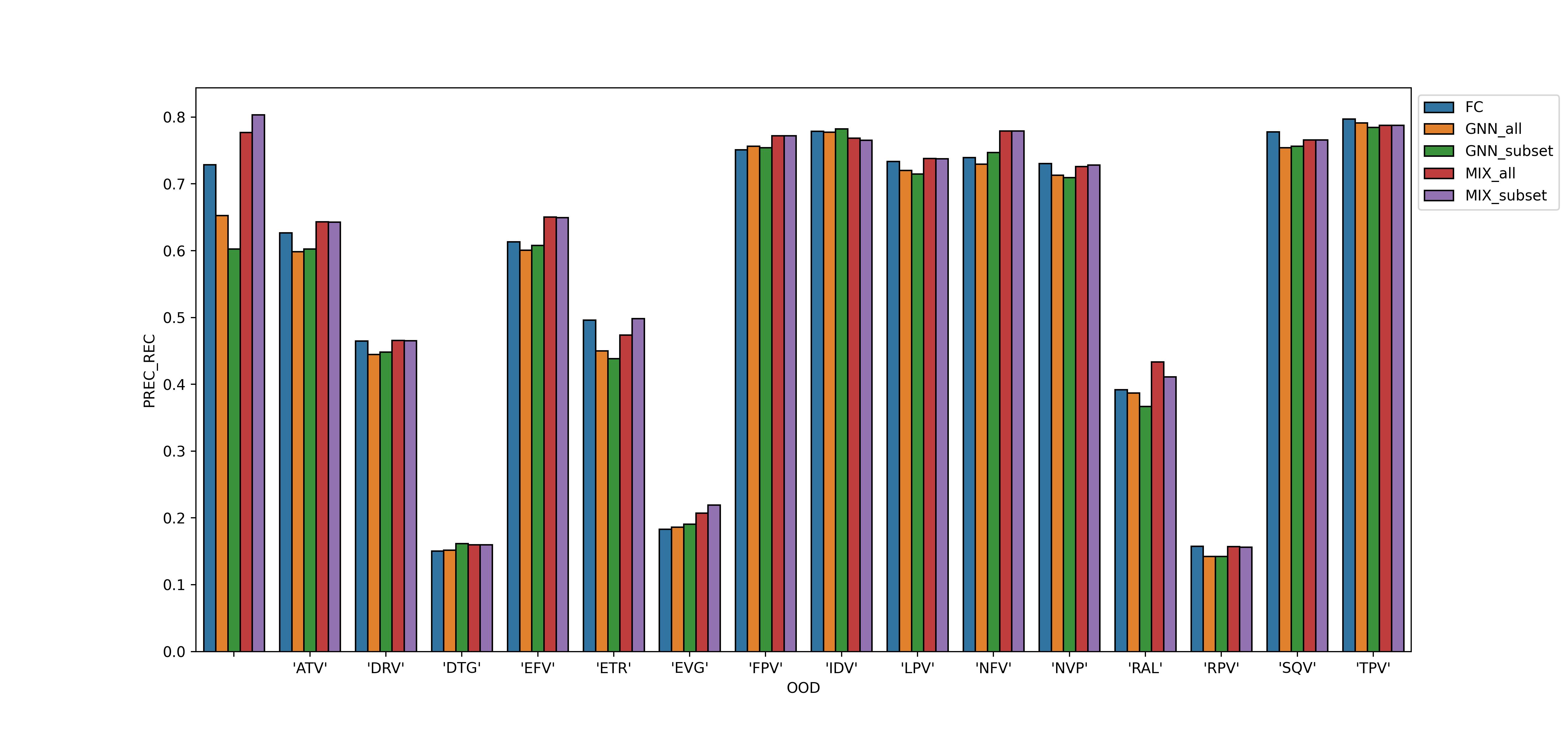}
    \caption{}
    \label{fig:sub2_prec_rec}
  \end{subfigure}
\caption{Panels (a), (b) and (c) represent the models' test performance in Accuracy, ROC AUC score and PREC-REC AUC score, respectively. The histograms are divided per OoD features. }
  \label{fig: graphs_results}
\end{figure*}

To facilitate visual comparison of model performance, the quantitative results presented in Table 1 are graphically depicted in Figure \ref{fig: graphs_results}

\begin{table}[ht]
\centering
\renewcommand{\arraystretch}{1.}

\begin{tabular*}{\columnwidth}{@{\extracolsep\fill}llll@{\extracolsep\fill}}
\toprule
    \# Therapies per patient & Label & \# Patients \\

\midrule
1	& 0	& 4659 \\
1	& 1	& 3363 \\
\hline
2	& 0	& 1781 \\
2	& 1	& 1230\\
\hline
3	& 0	& 691\\
3	& 1	& 501\\
\hline
4	& 0	& 253\\

4	& 1	& 235\\
\hline
5	& 0	& 96\\

5	& 1	& 118\\
\hline
6	& 0	& 44\\

6	& 1	& 49\\
\hline
7	& 0	& 19\\

7	& 1	& 26\\
\hline
8	& 0	& 14\\

8	& 1	& 14\\
\hline
9	& 0	& 4\\

9	& 1 & 	10\\
\hline
10	& 0	& 5\\
\hline
11	& 1	& 3\\

11	& 1	& 1\\
\hline
12	& 1	& 2\\
\hline
15	& 1	& 1\\

   \bottomrule
  \end{tabular*}
  \caption{Therapies count per patient}
  \label{tab:therapies per patient}
\end{table}

\section{Additional results}
In this section, we present additional computational results derived from the Treatment Change Episode (TCE) repository, which is part of the Stanford University HIV drug resistance database (\url{https://hivdb.stanford.edu/TCEs/}). The TCE repository contains over 1,500 TCE cases, each documented using an XML schema that encompasses essential clinical data, including: (i) previously used ARTs, (ii) baseline measurements of plasma HIV-1 RNA load, CD4 cell counts, and genotypic resistance test results, (iii) subsequent salvage therapies, and (iv) HIV-1 RNA levels monitored during salvage therapy.

A crucial pre-processing step was carried out to align the TCE data from Stanford with those from the EIDB, both in structure and content, to ensure consistency and compatibility for the analysis. We selected PTEs that fulfilled the criteria for evaluating therapeutic outcomes, as defined in the Standard Datum definition of the main paper. This process led to the identification of a more focused dataset of 562 PTEs.

Out of this dataset, only 63 PTEs (11.2\%) were categorized as successes. The dataset is notably imbalanced, with a significant skew toward treatment failures, contrary from real-world scenario. This can be attributed to two primary factors. First, the data collection process may not have prioritized creating a representative sample of HIV treatments. Clinicians contributing to the TCE repository may have simply provided genetic sequences that were readily available, often those related to treatment failures, as genotyping was typically used in such contexts for many years. Second, the data primarily reflects ARTs administered before 2011, prior to the widespread adoption of highly effective regimens, including second-generation INSTIs. As a result, the dataset is heavily weighted toward ARTs that had higher failure rates compared to more recent treatments, thus amplifying the prevalence of therapy failures.

The TCE dataset was utilized as an external validation set to test models pre-trained on EIDB-derived datasets, with results detailed in Table \ref{tab:results_performance_tce}.
It is worth noting, however, that not all the drugs categorized as OoD in the main paper are included in these results. This limitation arises because, when constructing test sets for each OoD drug as outlined in Section 3.4 of the main paper, the TCE dataset lacked sufficient samples for certain drugs where simulating the OoD scenario was possible with the EIDB dataset. While this dataset does not fully capture a real-world scenario—where successful cases far outnumber failures, and the types of therapies differ—the outcomes align with those observed using the EIDB dataset. Importantly, this analysis confirms that incorporating knowledge derived from Stanford scores significantly enhances prediction accuracy compared to using genotypic information in a standard tabular format.

\captionsetup[subtable]{labelformat=empty}

\begin{table*}[htbp]
\centering
\caption{ Test set performances (\%). The best and second-best results are in \textbf{bold} and \underline{underlined}, respectively. * The Wilcoxon test is significant at level 0.01, $\star$ at level 0.05.}
\label{tab:results_performance_tce}

\setlength{\tabcolsep}{3pt} 
\renewcommand{\arraystretch}{0.9} 

\begin{tabular}{cccc}
\begin{subtable}[t]{0.23\textwidth}
\centering
\caption{ DRV}
\scriptsize 
\adjustbox{max width=\linewidth}{
\begin{tabular}{lccc}
\toprule
\textbf{Model} & \textbf{Acc} & \textbf{ROC} & \textbf{PR} \\
\midrule
 FC & \begin{tabular}[c]{@{}l@{}}  67.9\\  ($\pm4.15$)  \end{tabular} & \begin{tabular}[c]{@{}l@{}}  \underline{53.9}\\ ($\pm5.99$)  \end{tabular} & \begin{tabular}[c]{@{}l@{}}  76.5\\ ($\pm5.49$)  \end{tabular} \\
 GNN$_A$ & \begin{tabular}[c]{@{}l@{}}  \underline{68.8}\\ ($\pm3.99$)  \end{tabular} & \begin{tabular}[c]{@{}l@{}} 52.7\\ ($\pm10.23$)  \end{tabular} & \begin{tabular}[c]{@{}l@{}}  \underline{92.6}\\ ($\pm3.20$)  \end{tabular} \\
 GNN$_S$ & \begin{tabular}[c]{@{}l@{}}  65.2\\ ($\pm4.86$)  \end{tabular} & \begin{tabular}[c]{@{}l@{}}  48.4\\ ($\pm9.25$)  \end{tabular} & \begin{tabular}[c]{@{}l@{}}  91.7\\ ($\pm3.39$)  \end{tabular} \\
 MIX$_A$ & \begin{tabular}[c]{@{}l@{}}  \textbf{69.6}\\ ($\pm4.61$)  \end{tabular} & \begin{tabular}[c]{@{}l@{}}  53.7\\ ($\pm9.05$)  \end{tabular} & \begin{tabular}[c]{@{}l@{}}  \textbf{93.2}\\ ($\pm2.76$)  \end{tabular} \\
 MIX$_S$ & \begin{tabular}[c]{@{}l@{}}  \underline{68.8}\\ ($\pm4.05$)  \end{tabular} & \begin{tabular}[c]{@{}l@{}}  \textbf{54.0}\\ ($\pm8.40$)  \end{tabular} & \begin{tabular}[c]{@{}l@{}}  \textbf{93.2}\\ ($\pm2.55$)  \end{tabular} \\
\bottomrule
\end{tabular}
}
\end{subtable}
&
\begin{subtable}[t]{0.23\textwidth}
\centering
\caption{ DTG}
\scriptsize
\adjustbox{max width=\linewidth}{
\begin{tabular}{lccc}
\toprule
\textbf{Model} & \textbf{Acc} & \textbf{ROC} & \textbf{PR} \\
\midrule
 FC & \begin{tabular}[c]{@{}l@{}} 65.2\\ ($\pm4.57$)  \end{tabular} & \begin{tabular}[c]{@{}l@{}}  53.0\\ ($\pm6.55$)  \end{tabular} & \begin{tabular}[c]{@{}l@{}} 76.0\\ ($\pm4.42$)  \end{tabular} \\
 GNN$_A$ & \begin{tabular}[c]{@{}l@{}}  \textbf{70.5}\\ ($\pm4.26$)  \end{tabular} & \begin{tabular}[c]{@{}l@{}}  50.0\\ ($\pm8.56$)  \end{tabular} & \begin{tabular}[c]{@{}l@{}}  92.6\\ ($\pm2.88$)  \end{tabular} \\
 GNN$_S$ & \begin{tabular}[c]{@{}l@{}}  65.2\\ ($\pm4.40$)  \end{tabular} & \begin{tabular}[c]{@{}l@{}}  \textbf{61.0}\\ ($\pm9.35$)  \end{tabular} & \begin{tabular}[c]{@{}l@{}}  \textbf{94.7}\\ ($\pm1.98$)  \end{tabular} \\
 MIX$_A$ & \begin{tabular}[c]{@{}l@{}}  67.0\\ ($\pm4.18$)  \end{tabular} & \begin{tabular}[c]{@{}l@{}}  53.3\\ ($\pm9.48$)  \end{tabular} & \begin{tabular}[c]{@{}l@{}}  93.3\\ ($\pm2.71$)  \end{tabular} \\
 MIX$_S$ & \begin{tabular}[c]{@{}l@{}}  \underline{67.9}\\ ($\pm4.31$)  \end{tabular} & \begin{tabular}[c]{@{}l@{}}  \underline{53.7}\\ ($\pm9.08$)  \end{tabular} & \begin{tabular}[c]{@{}l@{}}  \underline{93.4}\\ ($\pm2.51$)  \end{tabular} \\
\bottomrule
\end{tabular}
}
\end{subtable}
&
\begin{subtable}[t]{0.23\textwidth}
\centering
\caption{ EVG}
\scriptsize
\adjustbox{max width=\linewidth}{
\begin{tabular}{lccc}
\toprule
\textbf{Model} & \textbf{Acc} & \textbf{ROC} & \textbf{PR} \\
\midrule
 FC & \begin{tabular}[c]{@{}l@{}}  66.1\\ ($\pm4.45$)  \end{tabular} & \begin{tabular}[c]{@{}l@{}}  53.8\\ ($\pm5.78$)  \end{tabular} & \begin{tabular}[c]{@{}l@{}}  77.3\\ ($\pm4.77$)  \end{tabular} \\
 GNN$_A$ & \begin{tabular}[c]{@{}l@{}}  67.0\\ ($\pm4.66$)  \end{tabular} & \begin{tabular}[c]{@{}l@{}}  62.9\\ ($\pm7.74$)  \end{tabular} & \begin{tabular}[c]{@{}l@{}}  \textbf{95.3}\\ ($\pm1.62$)  \end{tabular} \\
 GNN$_S$ & \begin{tabular}[c]{@{}l@{}}  \textbf{89.3*}\\ ($\pm2.66$)  \end{tabular} & \begin{tabular}[c]{@{}l@{}}  49.7\\ ($\pm9.43$)  \end{tabular} & \begin{tabular}[c]{@{}l@{}}  91.3\\ ($\pm3.21$)  \end{tabular} \\
 MIX$_A$ & \begin{tabular}[c]{@{}l@{}} 

 75.9$\star$ \\ ($\pm4.49$)  \end{tabular} & \begin{tabular}[c]{@{}l@{}}  \underline{58.4}\\ ($\pm9.38$)  \end{tabular} & \begin{tabular}[c]{@{}l@{}}  \underline{94.2}\\ ($\pm2.38$)  \end{tabular} \\
 MIX$_S$ & \begin{tabular}[c]{@{}l@{}}  \underline{82.1*}\\ ($\pm3.48$)  \end{tabular} & \begin{tabular}[c]{@{}l@{}}  \textbf{58.7}\\ ($\pm9.81$)  \end{tabular} & \begin{tabular}[c]{@{}l@{}}  94.1\\ ($\pm2.24$)  \end{tabular} \\
\bottomrule
\end{tabular}
}
\end{subtable}
&
\begin{subtable}[t]{0.23\textwidth}
\centering
\caption{ FPV}
\scriptsize
\adjustbox{max width=\linewidth}{
\begin{tabular}{lccc}
\toprule
\textbf{Model} & \textbf{Acc} & \textbf{ROC} & \textbf{PR} \\
\midrule
 FC & \begin{tabular}[c]{@{}l@{}}  67.9\\ ($\pm4.33$)  \end{tabular} & \begin{tabular}[c]{@{}l@{}}  52.5\\ ($\pm5.51$)  \end{tabular} & \begin{tabular}[c]{@{}l@{}}  75.3\\ ($\pm5.49$)  \end{tabular} \\
 GNN$_A$ & \begin{tabular}[c]{@{}l@{}}  \underline{70.5}\\ ($\pm3.61$)  \end{tabular} & \begin{tabular}[c]{@{}l@{}}  \textbf{55.5}\\ ($\pm8.58$)  \end{tabular} & \begin{tabular}[c]{@{}l@{}}  \textbf{93.4}\\ ($\pm2.49$)  \end{tabular} \\
 GNN$_S$ & \begin{tabular}[c]{@{}l@{}}  68.8\\ ($\pm4.77$)  \end{tabular} & \begin{tabular}[c]{@{}l@{}}  \underline{55.0}\\ ($\pm8.40$)  \end{tabular} & \begin{tabular}[c]{@{}l@{}}  92.2\\ ($\pm3.72$)  \end{tabular} \\
 MIX$_A$ & \begin{tabular}[c]{@{}l@{}} \textbf{80.4}$\star$\\ ($\pm3.73$)  \end{tabular} & \begin{tabular}[c]{@{}l@{}}  50.8\\ ($\pm9.34$)  \end{tabular} & \begin{tabular}[c]{@{}l@{}}  92.8\\ ($\pm2.58$)  \end{tabular} \\
 MIX$_S$ & \begin{tabular}[c]{@{}l@{}} \textbf{80.4}$\star$\\ ($\pm3.24$)  \end{tabular} & \begin{tabular}[c]{@{}l@{}}  50.9\\ ($\pm10.1$)  \end{tabular} & \begin{tabular}[c]{@{}l@{}}  \underline{92.9}\\ ($\pm2.58$)  \end{tabular} \\
\bottomrule
\end{tabular}
}
\end{subtable}\\[-2mm] 

\begin{subtable}[t]{0.23\textwidth}
\centering
\caption{ RPV}
\scriptsize 
\adjustbox{max width=\linewidth}{
\begin{tabular}{lccc}
\toprule
\textbf{Model} & \textbf{Acc} & \textbf{ROC} & \textbf{PR} \\
\midrule
 FC & \begin{tabular}[c]{@{}l@{}}  67.0\\  ($\pm4.59$)  \end{tabular} & \begin{tabular}[c]{@{}l@{}}  52.9\\ ($\pm5.21$)  \end{tabular} & \begin{tabular}[c]{@{}l@{}}  76.6\\ ($\pm5.84$)  \end{tabular} \\
 GNN$_A$ & \begin{tabular}[c]{@{}l@{}}  68.8\\ ($\pm3.32$)  \end{tabular} & \begin{tabular}[c]{@{}l@{}} 53.9\\ ($\pm9.36$)  \end{tabular} & \begin{tabular}[c]{@{}l@{}}  93.5\\ ($\pm2.54$)  \end{tabular} \\
 GNN$_S$ & \begin{tabular}[c]{@{}l@{}}  66.1\\ ($\pm4.76$)  \end{tabular} & \begin{tabular}[c]{@{}l@{}}  \textbf{59.0}\\ ($\pm8.14$)  \end{tabular} & \begin{tabular}[c]{@{}l@{}}  94.4\\ ($\pm2.15$)  \end{tabular} \\
 MIX$_A$ & \begin{tabular}[c]{@{}l@{}}  \underline{69.6}\\ ($\pm3.90$)  \end{tabular} & \begin{tabular}[c]{@{}l@{}}  \underline{54.2}\\ ($\pm9.20$)  \end{tabular} & \begin{tabular}[c]{@{}l@{}}  93.4\\ ($\pm2.81$)  \end{tabular} \\
 MIX$_S$ & \begin{tabular}[c]{@{}l@{}}  \textbf{70.5}\\ ($\pm3.88$)  \end{tabular} & \begin{tabular}[c]{@{}l@{}}  \underline{54.2}\\ ($\pm9.91$)  \end{tabular} & \begin{tabular}[c]{@{}l@{}}  93.4\\ ($\pm2.84$)  \end{tabular} \\
\bottomrule
\end{tabular}
}
\end{subtable}
&
\begin{subtable}[t]{0.23\textwidth}
\centering
\caption{ TPV}
\scriptsize
\adjustbox{max width=\linewidth}{
\begin{tabular}{lccc}
\toprule
\textbf{Model} & \textbf{Acc} & \textbf{ROC} & \textbf{PR} \\
\midrule
 FC & \begin{tabular}[c]{@{}l@{}} 70.5\\ ($\pm3.51$)  \end{tabular} & \begin{tabular}[c]{@{}l@{}}  48.9\\ ($\pm5.32$)  \end{tabular} & \begin{tabular}[c]{@{}l@{}} 73.4\\ ($\pm6.14$)  \end{tabular} \\
 GNN$_A$ & \begin{tabular}[c]{@{}l@{}}  75.0\\ ($\pm4.20$)  \end{tabular} & \begin{tabular}[c]{@{}l@{}}  \textbf{60.6}\\ ($\pm8.64$)  \end{tabular} & \begin{tabular}[c]{@{}l@{}}  94.7\\ ($\pm2.07$)  \end{tabular} \\
 GNN$_S$ & \begin{tabular}[c]{@{}l@{}}  73.2\\ ($\pm4.15$)  \end{tabular} & \begin{tabular}[c]{@{}l@{}}  50.4\\ ($\pm8.71$)  \end{tabular} & \begin{tabular}[c]{@{}l@{}}  91.7\\ ($\pm3.63$)  \end{tabular} \\
 MIX$_A$ & \begin{tabular}[c]{@{}l@{}}  \underline{75.9}\\ ($\pm3.37$)  \end{tabular} & \begin{tabular}[c]{@{}l@{}}  \underline{50.5}\\ ($\pm8.72$)  \end{tabular} & \begin{tabular}[c]{@{}l@{}}  92.9\\ ($\pm2.47$)  \end{tabular} \\
 MIX$_S$ & \begin{tabular}[c]{@{}l@{}}  \textbf{77.7}\\ ($\pm3.49$)  \end{tabular} & \begin{tabular}[c]{@{}l@{}}  49.8\\ ($\pm8.88$)  \end{tabular} & \begin{tabular}[c]{@{}l@{}}  92.7\\ ($\pm2.76$)  \end{tabular} \\
\bottomrule
\end{tabular}
}
\end{subtable}
&
\begin{subtable}[t]{0.23\textwidth}
\centering
\caption{ No OoD}
\scriptsize
\adjustbox{max width=\linewidth}{
\begin{tabular}{lccc}
\toprule
\textbf{Model} & \textbf{Acc} & \textbf{ROC} & \textbf{PR} \\
\midrule
 FC & \begin{tabular}[c]{@{}l@{}}  66.1\\ ($\pm4.16$)  \end{tabular} & \begin{tabular}[c]{@{}l@{}}  \underline{51.0}\\ ($\pm6.57$)  \end{tabular} & \begin{tabular}[c]{@{}l@{}}  73.9\\ ($\pm6.17$)  \end{tabular} \\
 GNN$_A$ & \begin{tabular}[c]{@{}l@{}}  58.0\\ ($\pm4.33$)  \end{tabular} & \begin{tabular}[c]{@{}l@{}}  \textbf{60.5}\\ ($\pm8.24$)  \end{tabular} & \begin{tabular}[c]{@{}l@{}}  \textbf{93.8}\\ ($\pm2.69$)  \end{tabular} \\
 GNN$_S$ & \begin{tabular}[c]{@{}l@{}}  \underline{72.3}\\ ($\pm4.21$)  \end{tabular} & \begin{tabular}[c]{@{}l@{}}  48.3\\ ($\pm9.22$)  \end{tabular} & \begin{tabular}[c]{@{}l@{}}  91.4\\ ($\pm3.24$)  \end{tabular} \\
 MIX$_A$ & \begin{tabular}[c]{@{}l@{}} 

 \textbf{73.2}$\star$\\ ($\pm3.73$)  \end{tabular} & \begin{tabular}[c]{@{}l@{}}  42.5\\ ($\pm7.95$)  \end{tabular} & \begin{tabular}[c]{@{}l@{}}  \underline{91.7}\\ ($\pm2.84$)  \end{tabular} \\
 MIX$_S$ & \begin{tabular}[c]{@{}l@{}}  68.8\\ ($\pm4.15$)  \end{tabular} & \begin{tabular}[c]{@{}l@{}}  44.7\\ ($\pm9.83$)  \end{tabular} & \begin{tabular}[c]{@{}l@{}}  90.9\\ ($\pm2.91$)  \end{tabular} \\
\bottomrule
\end{tabular}
}
\end{subtable}
\end{tabular}
\end{table*}

\end{document}